\documentclass[twocolumn,twocolappendix]{aastex631}
\usepackage[whole]{bxcjkjatype} 
\usepackage{amsmath}	
\usepackage{bm}

\newcommand{\mr}[1]{\mathrm{#1}}

\defcitealias{Sugimura:2020aa}{Paper~I}
\defcitealias{Hosokawa:2016aa}{H16}

\shorttitle{Massive and Wide First-Star Binaries}
\shortauthors{Sugimura et al.}

\begin{document}

\title{Formation of Massive and Wide First-star Binaries in Radiation Hydrodynamics Simulations}

\author[0000-0001-7842-5488]{Kazuyuki Sugimura}
\affiliation{Faculty of Science, Hokkaido University, Sapporo, Hokkaido 060-0810, Japan}
\affiliation{The Hakubi Center for Advanced Research, Kyoto University, Sakyo, Kyoto 606-8501, Japan}
\affiliation{Department of Physics, Kyoto University, Sakyo, Kyoto 606-8502, Japan}
\email{sugimura@sci.hokudai.ac.jp}

\author[0000-0002-8125-4509]{Tomoaki Matsumoto}
\affiliation{Faculty of Sustainability Studies, Hosei University, Fujimi, Chiyoda, Tokyo 102-8160, Japan}

\author[0000-0003-3127-5982]{Takashi Hosokawa}
\affiliation{Department of Physics, Kyoto University, Sakyo, Kyoto 606-8502, Japan}

\author[0000-0002-4317-767X]{Shingo Hirano}
\affiliation{Department of Astronomy, School of Science, University of Tokyo, Tokyo 113-0033, Japan}

\author[0000-0001-5922-180X]{Kazuyuki Omukai}
\affiliation{Astronomical Institute, Graduate School of Science, Tohoku University, Aoba, Sendai 980-8578, Japan}



\begin{abstract}
We study the formation of Pop III stars by performing radiation
hydrodynamics simulations for three different initial clouds
extracted from cosmological hydrodynamics simulations. Starting from the cloud
collapse stage, we follow the growth of protostars by accretion for $\sim 10^5\,\mr{yr}$ until the radiative feedback from the protostars suppresses the accretion and the stellar properties are nearly fixed.  
We find that the Pop III stars form in massive and wide binaries/small-multiple stellar systems, with masses
$>30\,M_\odot$ and separations $>2000\,\mr{au}$. 
We also find
that the properties
of the final stellar system correlate with those of the initial clouds: the total
mass increases with the cloud-scale accretion rate, and the angular
momentum of the binary orbit matches that of the initial cloud.  While
the total mass of the system in our simulations is consistent with our previous single-star formation simulations, individual masses are lower
due to mass sharing, suggesting potential modification in the extent of feedback from Pop III stars in the subsequent evolution of the Universe.  
We also identify such systems as  
mini-binaries embedded in a wider outer multiple-star system, which could evolve into progenitors for observed gravitational
wave events.
\end{abstract}


\section{Introduction} 
\label{sec:intro}

First generation metal-free stars, known as Population III (Pop III)
stars, are believed to have appeared at redshifts of
$z\sim20\,\text{--}\,30$ as the first sources of light in the history of
Universe (e.g.,
\citealp{Couchman:1986aa,Tegmark:1997aa,Abel:2002aa,Bromm:2002aa}; see
also \citealp{Glover:2013aa, Greif:2015aa,Klessen:2023ui} for recent
review).  Their formation proceeds as follows: First, a small embryonic
protostar forms as a result of the gravitational collapse of clouds at
the center of minihalos ($10^{5}-10^{6}\,M_\odot$)
\citep[e.g.,][]{Omukai:1998aa,Yoshida:2008aa} and then it continues to
grow by accretion \citep[e.g.,][]{Omukai:2003ab,Tan:2004aa} until the
radiative feedback from the protostar eventually quenches the accretion
(e.g.,
\citealp{Omukai:2002aa,McKee:2008aa,Hosokawa:2011aa,Hosokawa:2016aa},
hereafter H16; \citealp{Stacy:2012aa}).

Previous numerical studies that followed the growth of embryonic Pop III
stars up to the end of the accretion (e.g.,
\citealt{Hosokawa:2011aa,Susa:2013aa,Stacy:2013aa,Susa:2014aa,Hirano:2014aa,Hirano:2015aa,Stacy:2016aa};
\citetalias{Hosokawa:2016aa}) demonstrate that the Pop III stars tend to
be more massive than predicted from the present-day Salpeter-like
initial mass function (IMF). This allows Pop III stars to play crucial
roles in driving the subsequent cosmic evolution through various
processes, such as the reionization of the intergalactic medium by
stellar radiation
\citep[e.g.,][]{Schaerer:2002aa,Wise:2012aa,Ricotti:2016aa}, the metal
enrichment of the pristine gas by supernovae
\citep[e.g.,][]{Woosley:2002aa,Chiaki:2018ab,Abe:2021vp}, and the
seeding of supermassive black hole (BH) progenitors
\citep[e.g.,][]{Alvarez:2009aa,Jeon:2012aa}.

Observations show that massive stars are predominantly in multiple
systems in the nearby Universe \citep[e.g.,][]{Duchene:2013aa}.  If this
is also the case in the early Universe, massive Pop III stars would also
be in multiples.  In this case, mass reservoir in the parental cloud
will be shared among forming stars and the individual stellar masses
will decrease.  Such Pop III binaries are attracting attention as they
may later evolve into binary black holes detectable by gravitational
waves (e.g.,
\citealp{Kinugawa:2014aa,Hartwig:2016aa,Abbott:2016aa,Tanikawa:2021aa,Liu:2021a},
but see also \citealp{Belczynski:2017aa}) or the first X-ray binaries,
which are important heating sources of the intergalactic medium and
should be constrained by future 21cm line observations
\citep[e.g.,][]{Mirabel:2011aa,Dewdney:2009aa}.  The X-ray background
can also change the mode of the Pop III star formation
\citep[e.g.,][]{Ricotti:2016ab,Park:2021aa,Park:2021ab,Park:2023ta}.

On the theoretical side, seminal works found that Pop III stars in fact
end up in multiple-star systems, by way of numerical simulations using
particle-based codes
\citep{Susa:2013aa,Stacy:2013aa,Susa:2014aa,Stacy:2016aa}.  In those
calculations, however, protostellar feedback may have been
underestimated because the density-dependent resolution of the
particle-based codes was insufficient to directly follow the propagation
of the ionizing radiation in the low-density polar regions around each
protostar \citep{Susa:2013aa}.

Later, in \cite{Sugimura:2020aa} (hereafter Paper I), we confirmed that
Pop III stars do form in a multiple-stellar system by performing a
simulation with a newly developed radiation hydrodynamics (RHD) code,
SFUMATO-RT, which employs the adaptive-mesh-refinement \citep[AMR;
][]{Berger:1989vo,Berger:1984ui} and the adaptive-ray-tracing
\citep[ART; ][]{Abel:2002ab} techniques to accurately follow the
ionizing radiation from multiple protostars.  In that work, the
formation of multiple protostars by gas fragmentation and their
long-term evolution are followed until the radiation feedback terminates
the accretion, thereby fixing the stellar properties.

Quite recently, other groups also performed similar simulations using
other AMR codes \citep{Latif:2022wf,Park:2023ta} and a moving-mesh code
\citep{Jaura:2022vn} with different radiation transfer modules. While
all groups agree on the formation of multiple stars, there is a
discrepancy on the effectiveness of radiation
feedback. \cite{Jaura:2022vn} argued that the trapping of ionizing
radiation near protostars significantly weakens the radiative feedback
unlike in the other simulations where efficient feedback is observed.
The cause of this discrepancy is still unclear and further investigation
is needed.

In \citetalias{Sugimura:2020aa}, our simulation was limited only to a
single case, despite the known diversity of the birth environments
\citep[e.g.,][]{Hirano:2014aa}.  We could not discuss such statistics as
the relation between the final stellar mass and the natal cloud
properties, found in 2D simulations \citep{Hirano:2014aa}.  Furthermore,
in \citetalias{Sugimura:2020aa}, we just focused on presenting the
overall evolution and the nature of resulting stellar system, leaving
analysis of detailed physical process to future work.

In this paper, we simulate Pop III star formation in three different
clouds extracted from cosmological simulations, using the same code as
in \citetalias{Sugimura:2020aa}.  Among the three clouds studied, two
are new cases while the one is the same as that already presented in
\citetalias{Sugimura:2020aa}.  Based on the results, we examine the
relation between the properties of the final stellar systems, such as
the total mass and binary separation, and those of the natal clouds.  We
also identify and describe in detail some interesting phenomena that can
play important roles in Pop III star formation.

The rest of this paper is organized as follows. In
Section~\ref{sec:methods}, we describe our numerical methods in
detail. In Section~\ref{sec:results}, we present our simulation results,
from which we then investigate relations between the stellar system and
natal clouds in Section~\ref{sec:relation}. In
Section~\ref{sec:discussion}, we discuss the role of radiative feedback
and the numerical resolution effects.  We conclude the paper in
Section~\ref{sec:conclusion}. The appendices provide a detailed
description of simulation methods and a supplementary analysis of
circum-stellar disks in our simulations.

\section{Numerical methods} 
\label{sec:methods}

\subsection{SFUMATO-RT} 
\label{sec:code}

Our radiation hydrodynamics code SFUMATO-RT has been developed to follow
the Pop III star formation and was first used in Paper I.  The code is
an extended version of an AMR self-gravitational magnetohydrodynamics
(MHD) code SFUMATO \citep{Matsumoto:2007aa,Matsumoto:2015ab}, with
addition of new modules to solve the non-equilibrium chemistry and
thermodynamics of primordial gas under the influence of radiation from
multiple protostars \citep[see][for other
extensions]{Sadanari:2021aa,Sadanari:2023vd,Kimura:2023}.

\subsubsection{Hydrodynamics} 
\label{sec:SFUAMTO}

We use SFUMATO's modules to solve hydrodynamics with self-gravity and
sink particles that represent accreting protostars.  We do not use the
MHD module since we ignore magnetic fields in this work.

The basic equations of hydrodynamics are
\begin{align}
\frac{\partial \rho}{\partial t} + \nabla  \cdot \left( \rho \bm{v} \right)=0\,,
\label{eq:1}\\
\rho \frac{\partial \bm{v}}{\partial t} + \rho \left( \bm{v} \cdot \nabla \right) \bm{v} = - \nabla P + \rho\,\bm{g}\,,
\label{eq:2}\\
\frac{\partial e}{\partial t} + \nabla \cdot \left[  \left(  e + P \right) \bm{v} \right] = \rho \bm{v} \cdot \bm{g} + \Gamma - \Lambda\,,
\label{eq:3}\\
P=(\gamma-1) e\,,
\label{eq:4}
\end{align}
where $\rho$ is the mass density, $\bm{v}$ the velocity, $P$ the
pressure, $\bm{g}$ the gravitational acceleration, $\Gamma$ the heating
rate, $\Lambda$ the cooling rate, and $\gamma$ the adiabatic exponent.
The mass density and pressure are related to the number density of
hydrogen atoms $n_\mathrm{H}$ as $\rho=n_\mathrm{H}\,\mu\,m_\mathrm{p}$
and $P=n_\mathrm{H}\,\sum_{\mr i}y(\mr{i})\,k_\mathrm{B}\,T$,
respectively, with $\mu$ the mean molecular weight per hydrogen atom,
$T$ the temperature, $m_\mathrm{p}$ the proton mass, $k_\mathrm{B}$ the
Boltzmann constant, and $y(\mr{i})$ the chemical abundances defined as
the abundance ratio of species i to the hydrogen nuclei.  We determine
$\Gamma$, $\Lambda$, $\gamma$, and $\mu$ from $T$ and $y(\mr{i})$
\citep[see, e.g.,][]{Omukai:1998aa}.  The chemical and thermal model is
described in Section~\ref{sec:chem_model}.  In this work, we use the
hydrodynamical scheme with second-order accuracy in space and time.

We consider both the self-gravity of the gas and the gravity by the sink
particles. The total gravitational acceleration is given by
 \begin{align}
  \bm{g} = - \nabla \Phi + \sum_\mathrm{i} \bm{g}_\mathrm{sink,i}\,,
 \label{eq:7}
 \end{align}
where $\Phi$ is the gravitational potential of the gas and
$\bm{g}_\mathrm{sink,i}$ is the gravitational acceleration due to the i-th sink
particle. Using the multigrid
solver, we obtain $\Phi$ from the Poisson equation,
 \begin{align}
 \nabla ^{2} \phi = 4 \pi G \rho\,,
 \label{eq:5}
 \end{align}
with Newton's gravitational constant $G$. As for
$\bm{g}_\mathrm{sink,i}$, we directly evaluate Newton's inverse-square
law,
 \begin{align}
 \bm{g}_\mathrm{sink,i}= - \frac{G\,M_\mathrm{i}
 (\bm{x}-\bm{x}_\mathrm{i})}{\left|\bm{x}-\bm{x}_\mathrm{i}\right|^3}\,,
 \label{eq:6}
 \end{align}
where $\bm{x}$ and $\bm{x}_\mathrm{i}$ are the positions of the gas and
the i-th particle, respectively.

Using the sink particle technique, we mask the neighborhoods of
protostars with extremely short timescales to perform long-term
simulations until the end of the accretion phase.  Below, we briefly
describe the implementation of sink particles in SFUMATO and refer the
reader to \cite{Matsumoto:2015ab} for details.

We create a new sink particle in a cell that satisfies the following
conditions \citep{Federrath:2010ab}:
\begin{itemize}
 \item[(i)] the density is higher than the sink threshold density
	    $n_\mathrm{sink}$.
 \item[(ii)] the cell is a local minimum of gravitational potential
	     $\Phi$.
 \item[(iii)] all the eigenvalues of the symmetric parts of the velocity
	      gradient tensor $\nabla_\mathrm{i}v_\mathrm{j}$ are
	      negative.
 \item[(iv)] the total energy of the gas within the sink radius
	     $r_\mathrm{sink}$ is negative
	     ($E_\mathrm{thermal}+E_\mathrm{kin}+E_\mathrm{grav} < 0$).
\end{itemize}

Around each sink particle, we define a virtual sphere with the radius
$r_\mathrm{sink}$ (also called the sink sphere), which is used for the
following purposes.  First, the sink particle accretes the gas from the
cells inside the sink sphere if the density exceeds $n_\mathrm{sink}$
until the excess disappears.  Second, the sink gravity is weakened
inside the sink sphere.  Finally, two sink particles merge when their
sink spheres overlap.  In our fiducial runs, we set
$r_\mathrm{sink}=64\,\mr{au}$ (16 times the minimum cell size) and
$n_\mathrm{sink}=2\times10^{11}\,\mathrm{cm^{-3}}$, as described in
Section~\ref{sec:sim_parm}.

The AMR technique allows us to follow fine structures near multiple
protostars at a relatively low computational cost. We refine the cells
so that the local Jeans length is resolved with at least $N_\mathrm{J}$
cells.  In this work, we set $N_J=16$, which is much higher than the
so-called Truelove condition $N_J\geq4$ \citep{Truelove:1997aa} for
avoiding artificial fragmentation.  Although it has recently been
claimed that higher resolution ($N_J\gtrsim 30$) is needed to follow the
turbulence amplification associated with gravitational collapse
\citep{Federrath:2011aa,Higashi:2021vt,Higashi:2022up}, we adopt the
above value to perform our simulations until the end of the accretion
phase.

\subsubsection{Non-equilibrium chemistry and thermodynamics}
\label{sec:chem_model}

One of major new additions of SFUMATO-RT to SFUMATO is a module for the
non-equilibrium chemistry and thermodynamics of the primordial gas,
which is essential to simulate Pop III star formation with realistic
thermal evolution.  We consider the chemical reactions among six
species, $\mathrm{H}^+$, $\mathrm{H}$, $\mathrm{H}_2$, $\mathrm{H}^-$,
$\mathrm{H}_2^+$, and $\mathrm{e}^{-}$, assuming that all the helium is
neutral.  To update $y(\mr{i})$ and $T$ consistently, we solve the
coupled equations for the chemical and temperature evolution, adopting
an implicit method for numerical stability.  Our chemical and thermal
models are summarized in Appendix~\ref{sec:chem_therm_models}.

In short, we consider all the chemical reactions and cooling and heating
processes that are relevant in the density range of $n_\mathrm{H}<
10^{13}\mr{cm^{-3}}$, as in \citetalias{Hosokawa:2016aa}.  Major
chemical reactions include the $\mathrm{H}^{-}$-channel $\mathrm{H}_2$
formation, the 3-body $\mathrm{H}_2$ formation, the collisional
dissociation and the photodissociation of $\mathrm{H}_2$, the
collisional ionization and the photoionization of $\mathrm{H}$, the
radiative recombination of $\mathrm{H}^{+}$. Major cooling and heating
processes include the $\mathrm{H}_2$ line cooling with line-trapping
effect, the $\mathrm{H}$ Ly$\alpha$ cooling, the $\mathrm{H}^-$
free-bound cooling, the $\mathrm{H}$ free-free cooling, the $\mathrm{H}$
free-bound cooling, the chemical cooling/heating, and the adiabatic
compression heating and expansion cooling.

Inside sink spheres, we solve the non-equilibrium chemistry and
thermodynamics as in the normal cells, but with neglected coupling to
the radiation to avoid artificial enhancement of radiative feedback.  We
adopt this approach to prevent potential numerical artifacts arising
from discontinuities between the interior and exterior of sink spheres.
The radiation transfer model is described in the next section.

\subsubsection{Radiation}
\label{sec:stellar_rad}

Another feature of SFUMATO-RT is addition of a radiation module to
handle the radiation from multiple sources.  We here consider three
types of radiation: extreme-ultraviolet radiation (EUV;
$h\nu>13.6\,\mathrm{eV}$), which photoionizes $\mr{H}$; far-ultraviolet
radiation (FUV; $11.2\,\mathrm{eV}<h\nu<13.6\,\mathrm{eV}$), which
photodissociates $\mr{H}_2$; and near-infrared radiation (NIR;
$h\nu<11.2\,\mathrm{eV}$), which photodetaches $\mr{H^-}$.  We solve the
radiation transfer for the EUV and FUV, while the NIR is assumed to be
optically thin, which is a good approximation in the accreting envelope
\citep[see][]{Hosokawa:2011aa}. Below, we describe the protostellar
model, the radiation transfer method, the sink-scale treatments, and the
coupling with chemistry and thermodynamics in this order.

\paragraph{Protostellar model}
\label{sec:method_protostellar}

We assign the radiative properties of Pop III protostars by tabulated
results of one-dimensional (proto)stellar evolution calculations under
constant accretion rates \citep{Hosokawa:2009aa,Hosokawa:2010aa}. The
table gives the luminosity $L_*$ and stellar radius $R_*$ for a given
set of the stellar mass $M$ and accretion rate $\dot{M}$. From $L_*$ and
$R_*$, we derive the effective temperature $T_\mr{eff}$, EUV emissivity
$S_\mr{EUV}$, and FUV emissivity $S_\mr{FUV}$, assuming a blackbody
spectrum (see Appendix~\ref{sec:proto_stellar_model} for more
details).  We also obtain the optically-thin $\mr{H^-}$ photodetachment
rate as a function of the distance from the protostar.

Following \citetalias{Hosokawa:2016aa}, we average the accretion rate
over 300 years to account for the timescale of gas transport through
unresolved parts of the accretion disk.
We only consider radiation from protostars more massive than
$M_\mr{rad,min}=5\,M_\odot$ to save computational cost since the UV
emissivities before the onset of the Kelvin-Helmholtz contraction are
extremely small (see Appendix~\ref{sec:proto_stellar_model}).

Our protostellar radiation model neglects the dependence of stellar
properties on the accretion history.  \citetalias{Hosokawa:2016aa} found
a case in which the intermittent plunge of fragments causes accretion
bursts, and thus the accretion history matters indeed.  In our
simulations, however, we expect the dependence on the accretion history
to be less important because the gas is mostly supplied to protostars by
continuous gas accretion from the circum-stellar disks, with modest
modulation due to binary interactions, except in early phases when
radiation feedback is negligible.

\paragraph{Radiative transfer method}
\label{sec:method_RT}

To solve the transfer of EUV and FUV radiation, we trace rays from each
source by way of the ART method
\citep{Abel:2002ab,Krumholz:2007ab,Wise:2011aa,Rosen:2017aa,Kim:2017aa}.
Below, we briefly explain our radiative transfer method and refer the
readers to the literature above for details about the ART method.  We
present our specific implementation to SFUMATO-RT in
Appendix~\ref{sec:ART}.

For the EUV radiation, we trace the direct photons from each protostar
considering the attenuation due to photoionization, with the diffuse
recombination photons treated by the on-the-spot approximation,
following \citetalias{Hosokawa:2016aa}.  At a distance $r$ from a
radiation source, the optical depth is obtained by the integration along
a ray as
\begin{align}
 \tau_\mr{EUV} = \int_{r_\mr{sink}}^r \sigma_\mr{pi}\,n_\mr{H}\,y(\mr{H}) \mr{d}r'\,,
\label{eq:9}
\end{align}
with $\sigma_\mr{pi}$ the effective photoionization cross-section
depending on the source's spectrum.  We omit the attenuation within the
sink sphere in Eq.~\eqref{eq:9} by setting the lower bound of the
integration to $r_\mr{sink}$, and separately consider the sink-scale
attenuation, as described below. With $\tau_\mr{EUV}$ along each ray, we
evaluate the EUV photoionization rate in each cell as
\begin{align}
 k_\mr{pi} = \sigma_\mr{pi}\,\frac{S_\mathrm{EUV}}{4\pi\,r^2}\,\langle e^{-\tau_\mr{EUV}}\rangle_\mr{cell}\,,
\label{eq:11}
\end{align}
and the EUV photo-heating rate as
\begin{align}
 \Gamma_\mr{pi} = \sigma_\mr{pi}\,\epsilon_\mr{EUV}\,\frac{S_\mathrm{EUV}}{4\pi\,r^2}\,\langle e^{-\tau_\mr{EUV}}\rangle_\mr{cell}\,,
\label{eq:10}
\end{align}
where $\epsilon_\mr{EUV} = \langle h\nu-13.6\mr{eV}\rangle$ is the mean
energy deposited in the gas per photoionization. The bracket $\langle\
\rangle_\mr{cell}$ denotes the average over the rays penetrating the
cell \citep[e.g., ][]{Wise:2011aa}.  If rays from multiple sources reach
the cell, we take the sum of each source's contribution.

For the FUV radiation, instead of the optical depth, we evaluate the
$\mr{H_2}$ column density
\begin{align}
N_\mr{H_2} = \int_{r_\mr{sink}}^r n_\mr{H}\,y(\mr{H_2}) \mr{d}r'\,,
\label{eq:8}
\end{align}
along a ray, to model the self-shielding effect.  We then estimate the
FUV photodissociation rate as
\begin{align}
 k_\mr{pd} = \sigma_\mr{pd}\,\frac{S_\mathrm{FUV}}{4\pi\,r^2}\,\langle f_\mr{shield}\rangle_\mr{cell}\,,
\end{align}
with the effective photodissociation cross-section $\sigma_\mr{pi}$ and
the self-shielding factor $f_\mr{shield}$.  For this work, we adopt the
widely-used formula
$f_\mr{shield}=\min[1,(N_\mr{H_2}/10^{14}\mr{cm^{-2}})^{-0.75}]$ from
\cite{Draine:1996aa}, although more elaborate ones have recently been
proposed \cite[e.g.,][]{Wolcott-Green:2019vl}.

In the ART framework, to achieve desired directional resolution, we
split rays hierarchically using the HEALPix library
\citep{Gorski:2005aa}.  At each ray level $l_\mr{ray}$, the spherical
$4\pi$ solid angle is sampled by $N_\mr{ray}=12\times 4^{l_\mr{ray}}$
rays that cover equal-area pixels with $4\pi/N_\mr{ray}$.  We start ray
tracing from each sink particle by injecting $N_\mr{ray,ini}=12\times
4^{l_\mr{ray,ini}}$ rays at the initial ray level ${l_\mr{ray,ini}}=3$.
While tracing the rays, we split a parent ray into four daughter rays at
one higher HEALPix level to ensure that at least $N_\mr{ray,min}$ rays
pass through each cell surface.  In this study, we use
$N_\mr{ray,min}=3$, following the resolution study of
\cite{Krumholz:2007ab}.  We have examined the effect of this choice by
performing a test run with $N_\mr{ray,min}=5$ and have confirmed that
its effect on the total mass evolution is insignificant, at least until
middle phases of star formation.  We randomly rotate the HEALPix
orientation at each ART step to mitigate artifacts due to insufficient
discretization \citep{Krumholz:2007ab}.

\paragraph{Sink-scale treatments}
\label{sec:method_sinkscale}

We need separate treatments for radiation rays inside sink spheres,
where the gas distribution is not resolved and can be artificial.  For
example, the inner geometrically thin part of an accretion disk in
reality may be artificially thick in simulations due to the softening of
the protostar's gravity inside the sink sphere.

To account for the shielding of central radiation by unresolved parts of
a disk inside a sink sphere, we allow only rays with elevation angles
greater than the disk thickness to cross the sink sphere.  Here, we
measure the disk thickness at the sink radius, assuming that the aspect
ratio is an increasing function of the radius. \footnote{It scales as
$H/R=(c_\mathrm{s}/\Omega)/R\sim R^{1/2}$ in a vertically-hydrostatic
isothermal disk} In practice, when a ray crosses a sink sphere, we check
whether the density is higher than the disk threshold density
$n_\mr{disk}=10^{-2}\,n_\mr{sink}$.  If this is the case, assuming that
the ray is traveling in the direction shielded by the disk, we terminate
the ray before it has any effect on the surrounding gas. We have
confirmed that the effect of radiative feedback is not sensitive to the
particular choice of $n_\mr{disk}$, by comparing axisymmetric test
simulations of accreting protostars with fiducial $n_\mr{disk}$ and 10
times larger $n_\mr{disk}$.

In addition to the disk shielding, we model the EUV absorption in the
polar directions following \citetalias{Hosokawa:2016aa}.  For each ray
emitted from a protostar, we measure the the density of the cell where
the ray crosses the sink sphere and check whether the Str\"omgren radius
$r_\mathrm{strm}$ for a homogeneous medium with this density (the
temperature is assumed to be $T=4\times10^4\,\mr{K}$, which is a typical
value for HII regions around Pop III stars) is smaller than the sink
radius. If this is the case, we assume that the EUV is consumed inside
the sink sphere and set a very large $\tau_\mr{EUV}$ (effectively no EUV
effect remains). Of course, the assumption of a homogeneous medium
inside the sink sphere is not very accurate: the density along the ray
increases inward in the case of a spherical free-falling flow as
$\propto r^{-3/2}$; conversely, the density may decrease in the case
that the polar regions are cleared due to the centrifugal force.

Alternatively, \cite{Jaura:2022vn} distributed photons at the center of
sink spheres and solve the radiation transfer inside those spheres.  As
a result, the effect of radiative feedback is strongly suppressed in
their simulations, as all photons are absorbed by the thick disks inside
the sink spheres. At present, the precise gas distributions surrounding
protostars are unknown, introducing uncertainties into the sink-scale
radiation model used in star formation simulations. We expect that
high-resolution RHD simulations, capable of resolving the star-disk
interfaces and the inner part of accretion disks
\citep[see][]{Kimura:2023}, are valuable in addressing this matter.

\paragraph{Coupling with chemistry}
\label{sec:method_ARTchem}

In each ART step, we trace rays on the fixed background of the density
and chemical composition.  We then update the chemistry and
thermodynamics under the fixed radiation field reconstructed from the
result of the previous ray tracing.  We perform the ART plus
chemistry/thermodynamics sub-cycles $n_\mr{sub}$ times per hydrodynamics
step.  In this study, we choose $n_\mr{sub}=2$ so that the maximum
propagation speed of the dissociation and ionization fronts in the
sub-cycles (two cells per one hydrodynamics step) is faster than that
associated with the hydrodynamic flows (one cell per one hydrodynamics
step).  As a result, we expect that the fronts can (in principle) reach
the positions determined by the ionization/recombination and
dissociation/formation equilibria, respectively.

\subsection{Initial conditions}
\label{sec:IC}

\begin{table*}
 \centering
 \caption{Summary of initial clouds.}
 \label{tab:cloud}
 \begin{tabular}{ccccccc} \hline \hline
Cloud$^{a}$ & $z_\mathrm{col}$ & $M_\mathrm{DM,halo}\, [M_\odot]$ & 
$R_\mathrm{cloud}^{b}\, [\mr{pc}]$ &
$M_\mathrm{cloud}^{c}\, [M_\odot]$ & 
$\dot{M}_\mathrm{cloud}^{d}\, [M_\odot/\mathrm{yr}]$ & $\lambda_\mathrm{cloud}^{e}$\\
\hline
Low-$\dot{M}$         & 18 & $4.2\times 10^5$ & $0.08$ &$130$ & $7.6\times 10^{-4}$ & $0.24$\\
Intermediate-$\dot{M}$& 24 & $2.5\times 10^5$ & $0.32$ & $860$ & $2.2\times 10^{-3}$ & $0.22$\\
High-$\dot{M}$        & 16 & $8.3\times 10^5$ & $0.25$ & $970$ & $5.2\times 10^{-3}$ & $0.14$\\
\hline
 \end{tabular}\\
 \begin{flushleft}
Notes. 
Column 1: cloud name, 
Column 2: collapse redshift,
Column 3: dark matter mass of the minihalo,
Column 4: cloud radius,
Column 5: gas mass of the cloud,
Column 6: cloud-scale accretion rate,
Column 7: cloud spin parameter.\\

$^{a}$ The High-$\dot{M}$, Intermediate-$\dot{M}$, and Low-$\dot{M}$
clouds in this paper are identical to the clouds in Cases A, C, and D in
\citetalias{Hosokawa:2016aa}, respectively. See \cite{Hirano:2015aa} and
\citetalias{Hosokawa:2016aa} for further information about the cloud
properties. The Intermediate-$\dot{M}$ cloud also corresponds to the
cloud studied in \citetalias{Sugimura:2020aa}.
All cloud parameters are evaluated when the central density reaches $n_\mr{H}=10^7\,\mr{cm^{-3}}$.\\
$^{b}$ The radius at
which the ratio of the enclosed mass to the local Jeans mass takes its
maximum value \citep[see][]{Hirano:2015aa}.
\\
$^{c}$ The gas mass within $R_\mr{cloud}$.
\\
$^{d}$ The accretion rate through the spherical surface at $R_\mr{cloud}$\\
$^{e}$ The spin parameter
 defined as $\lambda_\mr{cloud}=J_\mr{cloud}/\sqrt{2\,G\,M_\mr{cloud}^3\,R_\mr{cloud}}$,
with $J_\mr{cloud}$ the angular momentum within $R_\mr{cloud}$ \citep[see][]{Hirano:2015aa}.
 \end{flushleft}
\end{table*}

\begin{figure*}
 \centering
\includegraphics[width=17cm]{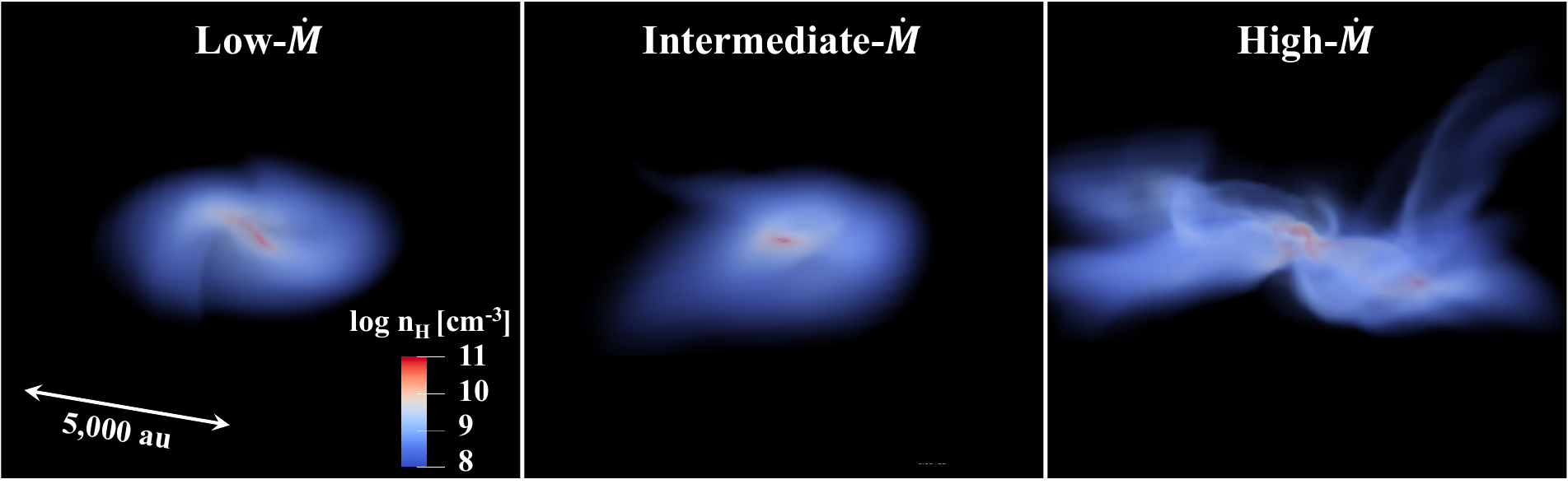}
\caption{The state of the cloud cores just before the formation of the
first protostars.  We show the density rendered image for the
Low-$\dot{M}$ (left), Intermediate-$\dot{M}$ (middle), and
High-$\dot{M}$ (right) clouds. } \label{fig:collapse_3d}
\end{figure*}

\begin{figure}
 \centering
\includegraphics[width=7cm]{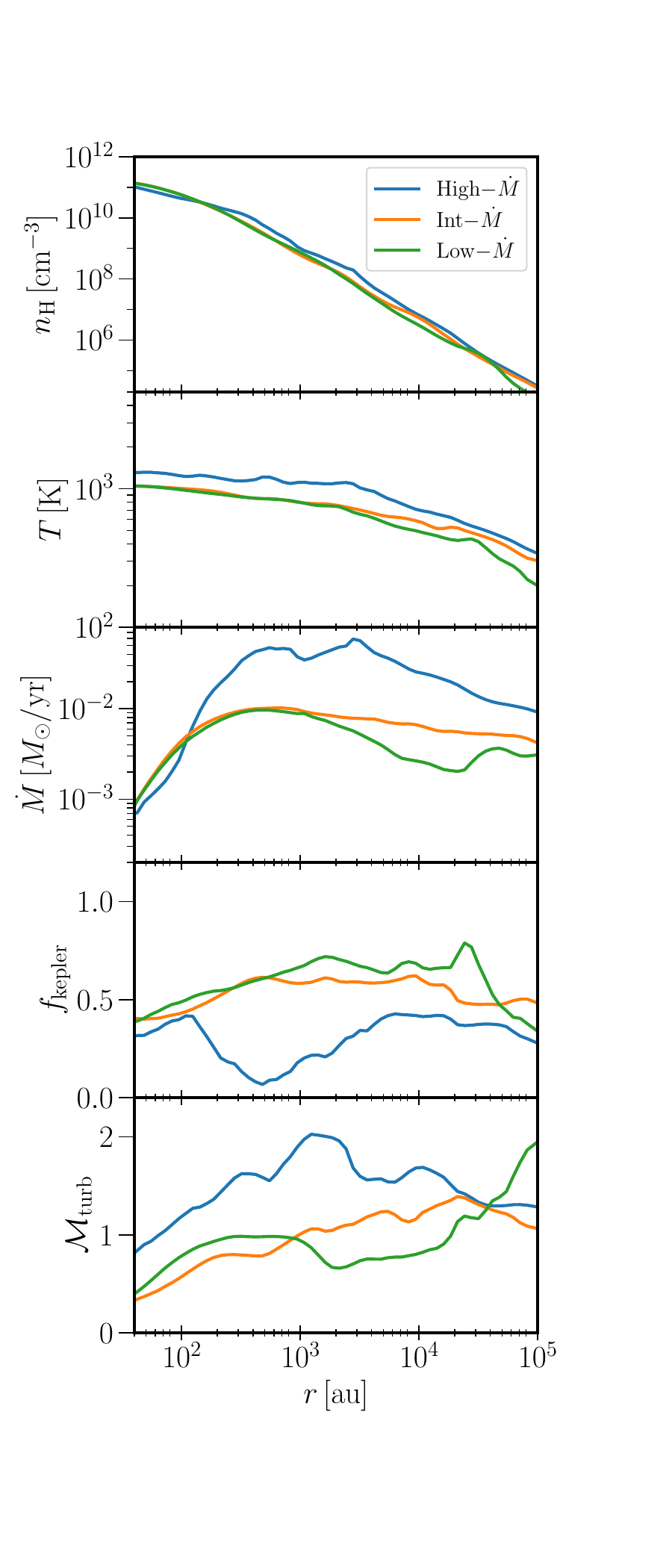}
\caption{The radial profiles of physical quantities just before the
first protostar formation for the three clouds (Low-$\dot{M}$, green
dotted; Intermediate-$\dot{M}$, orange dashed; and High-$\dot{M}$, blue
solid).  From top to bottom, the density $n_\mr{H}$, temperature $T$,
inflow rate $\dot{M}$, degree of rotational support $f_\mr{Kepler}$, and
turbulent Mach number $\mathcal{M}_\mr{turb}$ are plotted.  }
\label{fig:prof1d_collapse_comparison}
\end{figure}

We extract primordial star-forming clouds from cosmological 3D
SPH/N-body simulations \citep{Hirano:2014aa,Hirano:2015aa} as the
initial conditions for our 3D AMR RHD star-formation simulations.
Specifically, we remap particle-based simulation data of primordial
clouds early in the collapse phase (when the central density reaches
$10^6\,\mathrm{cm^{-3}}$) to Cartesian grids to generate the initial
conditions.

Table~\ref{tab:cloud} summarizes the properties of the three clouds
studied in this paper.  These clouds have a range of initial cloud-scale
accretion rate $\dot{M}_\mathrm{cloud}$ spanning from $10^{-3}$ to
$10^{-2}\,M_\odot/\mr{yr}$.  We give names to these clouds based on
their respective cloud-scale accretion rates: High-, Intermediate-, and
Low-$\dot{M}$ clouds. The Intermediate-$\dot{M}$ cloud is the same one
studied in Paper I.  Our selection of clouds with different
$\dot{M}_\mathrm{cloud}$ enables us to study a wide variety of Pop III
star forming environments, as $\dot{M}_\mathrm{cloud}$ is known to
correlate with the final stellar mass (\citealt{Hirano:2014aa},
\citetalias{Hosokawa:2016aa}).

After the onset of our re-simulation, the pre-stellar collapse continues
until the first protostar, i.e., sink particle, appears around the cloud
center.  Here, we show the state of the clouds just before the first
protostar formation by the 3D rendering of the central cores in
Fig.~\ref{fig:collapse_3d} and the 1D radial profiles of the entire
clouds in Fig.~\ref{fig:prof1d_collapse_comparison}.

Fig.~\ref{fig:collapse_3d} depicts the diverse morphology of central
cores in the three clouds.  The cores in the Low- and
Intermediate-$\dot{M}$ cases are rotating and have disk-like shapes,
with the former having a noticeable bar-spiral structure.  In contrast,
the core in the High-$\dot{M}$ case is turbulent and filamentary in
shape.  Those cores can be considered as the initial states of Pop III
star formation via accretion, as described in Section~\ref{sec:results}.

As seen in the density and temperature profiles shown in the top two
panels of Fig.~\ref{fig:prof1d_collapse_comparison}, the clouds undergo
so-called the runaway collapse, with a slightly increasing temperature
characterized by an effective polytropic index $\gamma_\mr{eff}\approx
1.1$ \citep{Omukai:1998aa}.  The third panel presents the inflow rates,
$\dot{M}$, at a given radius, which are overall consistent with the
values of $\dot{M}_\mathrm{cloud}$ (Table~\ref{tab:cloud}). Note,
however, that those two quantities do not exactly match at either radius
because the measurements are taken at different epochs while $\dot{M}$
gradually increases over time \citep{Hirano:2015aa}.  The bottom two
panels show the degree of rotational support
$f_\mr{kep}=\Omega/\Omega_\mr{Kepler}$ and the turbulent Mach number
$\mathcal{M}_\mr{turb}=v_\mr{turb}/c_\mr{s}$, where $\Omega_\mr{Kepler}$
is the Keplerian angular velocity and $v_\mr{turb}$ is the turbulent
velocity. The High-$\dot{M}$ case exhibits stronger turbulence than the
two other cases, with a lower degree of rotational support.  The initial
states of the clouds before the accretion phase are closely linked to
the properties of final Pop III systems, as discussed in
Section~\ref{sec:relation}.

\subsection{Simulation setup}
\label{sec:sim_parm}

\begin{table*}
 \centering
 \caption{Summary of fiducial runs.}
 \label{tab:fid}
\hspace*{-3cm}
 \begin{tabular}{ccccccccccc} \hline\hline
Run$^a$ & 
$L_\mathrm{box}\, [\mathrm{au}]$ & 
$\Delta x_\mr{min}\, [\mathrm{au}]$ & 
$r_\mathrm{sink}\, [\mathrm{au}]$ & 
$\Delta t_\mr{end}\, [\mathrm{yr}]$ & 
$N_\mathrm{sink}$  & 
$M_\mathrm{tot}\, [M_\odot]$ &  
$M_1\, [M_\odot]$ & 
$M_2\, [M_\odot]$ & 
$a_{12}\, [\mathrm{au}]$ & 
$M_{\mathrm{H16}}\, [M_\odot]$
\\
\hline
Low & $2.6\times10^5$ & $4$ & $64$  & $10^5$ & $2$ & $98$ & $64$ & $34$ & $8\times10^3$ & $67$      \\
Int & $5.2\times10^5$ & $4$ & $64$  & $10^5$ & $4$ & $148$ & $67$ & $56$ & $2\times10^4$ & $286$        \\
High & $5.2\times10^5$ & $4$ & $64$ & $8\times10^4$ & $2$ & $497$ & $370$ & $127$ & $2\times10^3$ & $>462^\mr{b}$  \\
\hline
 \end{tabular}\\
 \begin{flushleft}
Notes. Column 1:  run name,
Column 2: side length of the simulation box, 
Column 3: minimum cell size, 
Column 4: sink radius, 
Column 5: simulation duration since the first protostar formation, 
Column 6: number of sink particles surviving until the end of the simulation,
Column 7: total mass of the stars,
Column 8: mass of the most massive star,
Column 9: mass of the second most massive star,
Column 10: semi-major axis of the orbit of the most and the second most massive stars,
Column 11: stellar mass in \citetalias{Hosokawa:2016aa}.\\
$^\mr{a}$ - Low, Int, and High runs are for the High-$\dot{M}$, Intermediate-$\dot{M}$, Low-$\dot{M}$ clouds in Table~\ref{tab:cloud}, respectively.\\
$^\mr{b}$ - Accretion continues when \citetalias{Hosokawa:2016aa} stops their simulation.
 \end{flushleft}
\end{table*}

We perform three runs, namely, High, Int, and Low runs, corresponding to
the High-$\dot{M}$, Intermediate-$\dot{M}$, and Low-$\dot{M}$ clouds in
Section~\ref{sec:IC}, respectively. Table~\ref{tab:fid} summarizes our
simulation setup and results.

We set the side length of our simulation box to $2.6$ or
$5.2\,\mathrm{pc}$ to encompass the collapsing region of the cloud with
sufficient margin.  The minimum cell size at the highest AMR level is
$\Delta x_\mr{min} = 4\,\mathrm{au}$, with the sink radius of
$r_\mathrm{sink}=64\,\mathrm{au}$, equivalent to 16 times $\Delta
x_\mr{min}$.  Such a large number of cells per sink radius is necessary
to resolve geometrically-thin accretion disks with the thickness of
$\sim10\,\mr{au}$ around the sink radius: inadequate resolution would
artificially align the disk to one of the Cartesian axes since the gas
can be advected only through the cell surfaces in grid simulations.  We
adopt a sink threshold density of
$n_\mathrm{sink}=2\times10^{11}\,\mathrm{cm^{-3}}$, ensuring that the
corresponding Jeans length matches the sink diameter of
$2\,r_\mathrm{sink}$ (with $T=10^3\,\mathrm{K}$, representing a typical
temperature of neutral gas at density $n_\mathrm{sink}$).  We terminate
our simulations at $\Delta t = 80,000\,\mr{yr}$ or $100,000\,\mr{yr}$,
when accretion is quenched by radiation feedback and the total mass is
almost fixed.
 \section{Numerical results} 
 \label{sec:results}

In this section, we present our simulation results.  We first describe
the overall evolution in Section~\ref{sec:overall}, and then examine
noteworthy interesting phenomena in detail in
Section~\ref{sec:characteristic}.

 \subsection{Overall evolution} 
\label{sec:overall}

\begin{figure*}
 \centering \hspace*{-0.5cm}
\includegraphics[width=19cm]{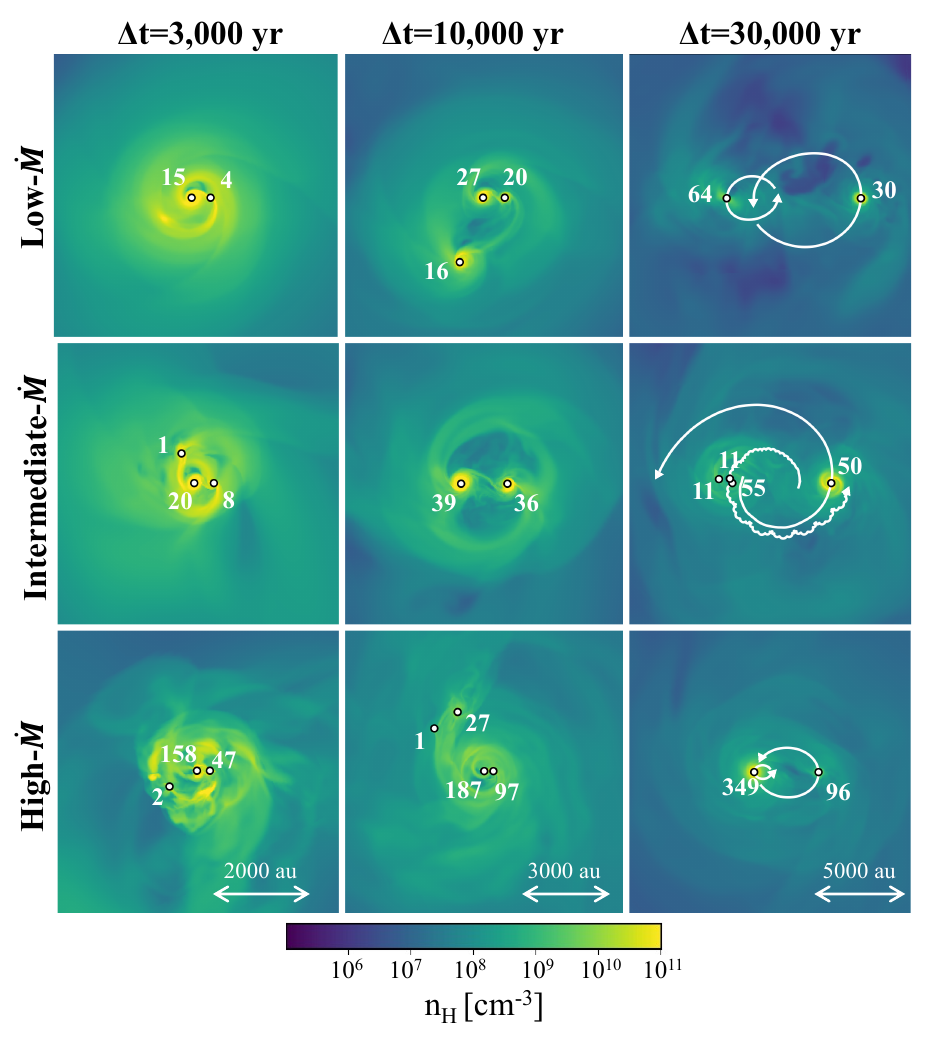}
\caption{Face-on slice density in the Low-$\dot{M}$ (top),
Intermediate-$\dot{M}$ (middle), and High-$\dot{M}$ (bottom) cases at
$\Delta t =3,000$ (left), $10,000$ (middle), and $30,000\,\mathrm{yr}$
(right) since the first protostar formation. The projected positions of
the sink particles are shown as open circles, with their masses
indicated in $M_\odot$.  Note that the size of the sink particles shown
is larger than the actual size for better visibility.  In the right
column, we show the trajectories of the most and second most massive
protostars.  } \label{fig:evolution_faceon_2d}
\end{figure*}

\begin{figure*}
 \centering \hspace*{-0.5cm}
\includegraphics[width=19cm]{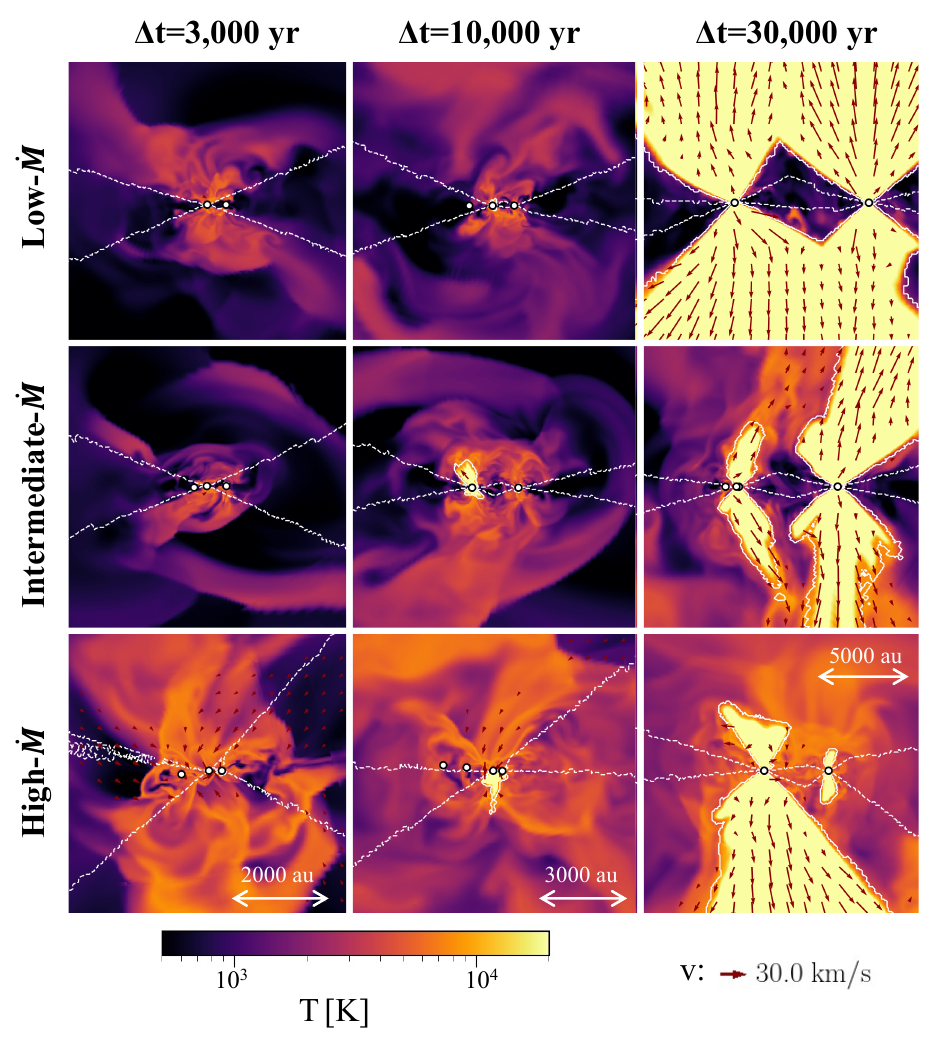}
\caption{Same as Fig.~\ref{fig:evolution_faceon_2d} but for the edge-on
slice temperature. The bipolar photoionization and photodissociation
fronts are demarcated with solid and dashed lines. The gas velocity is
indicated by arrows whose length is proportional to its amplitude.}
\label{fig:evolution_edgeon_2d}
\end{figure*}

First, we provide an overview of the star formation process in the three
clouds simulated in this paper.  To visually depict the evolution, we
show the face-on slice density in Fig.~\ref{fig:evolution_faceon_2d} and
the edge-on slice temperature in Fig.~\ref{fig:evolution_edgeon_2d} for
the Low- (top), Intermediate- (middle), and High-$\dot{M}$ (bottom) runs
at different evolutionary stages.  Specifically, we plot the snapshots
at $\Delta t =3,000\,\mathrm{yr}$ (left), $10,000\,\mathrm{yr}$
(middle), and $30,000\,\mathrm{yr}$ (right), with $\Delta t$ being the
time since the first protostar formation.  In all three cases, multiple
protostars are formed as a result of the initial disk fragmentation
(Fig.~\ref{fig:evolution_faceon_2d}, left), in agreement with previous
simulations of the Pop III star formation that followed the early phase
\citep[e.g.,][]{Machida:2008aa,Clark:2011ab,Greif:2011aa,Greif:2012aa,Smith:2011aa,Hirano:2017aa,Susa:2019wj,Sharda:2019aa}.
Subsequently, the protostars grow in mass via gas accretion
(Fig.~\ref{fig:evolution_faceon_2d}, middle) until the radiation
feedback quenches it (Fig.~\ref{fig:evolution_edgeon_2d}, right).  The
resulting stellar system is either a binary of two stars (in the Low-
and High-$\dot{M}$ cases) or a binary of a single star and a
mini-triplet system (in the Intermediate-$\dot{M}$ case), as shown in
Fig.~\ref{fig:evolution_faceon_2d} (right).

\begin{figure*}
 \centering \hspace*{-1cm}
\includegraphics[width=20cm]{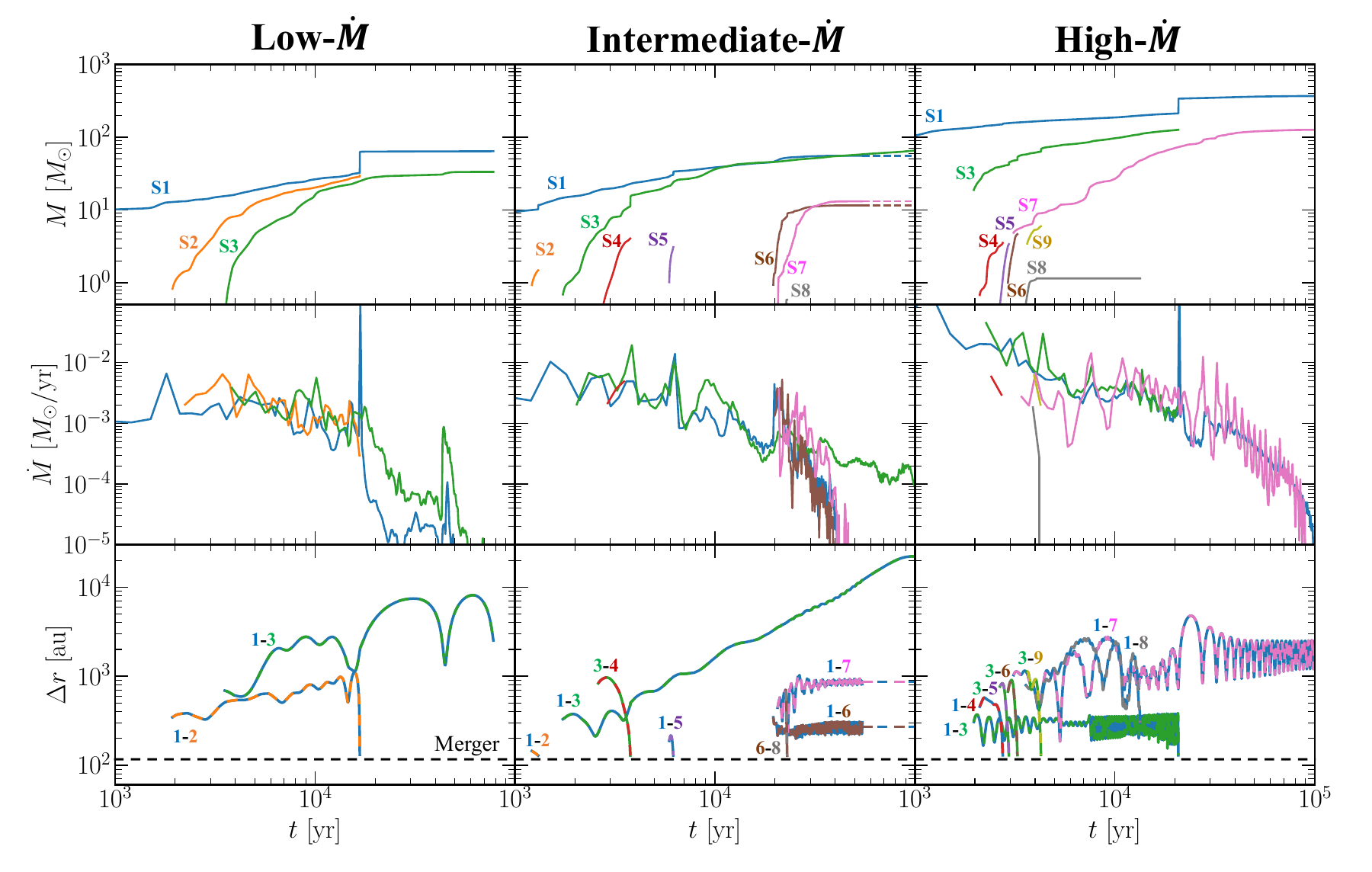} \caption{
Evolution of protostars, i.e., sink particles, in the three clouds: from
left to right columns, Low-, Intermediate- and High-$\dot{M}$ (right)
clouds. From top to bottom, the masses, accretion rates (averaged over
$300\,\mr{yr}$), and distances between selected pairs are plotted. In
the top two panels, the same color is used for the same protostar, whose
ID is indicated in the top panel, while a combination of the two colors
of the member stars is used to indicate the pair in the bottom panel.
The black dashed line in the bottom panel indicates the distance below
which two sink particles are assumed to merge. In the
Intermediate-$\dot{M}$ case, we do not follow the individual dynamics of
the mini-triplet after $t = 5.5 \times 10^4 \,\mr{yr}$ (dashed, see
\citetalias{Sugimura:2020aa}).  In the High-$\dot{M}$ case, the second
sink particle does not appear in the plot as it merges with the first
one at $\Delta t = 100\,\mr{yr}$, before the beginning of the plot. }
\label{fig:sink_detail}
\end{figure*}

To see the evolution, we plot the mass (top), accretion rate (middle),
and separation of selected pairs (bottom) of protostars, i.e., sink
particles, in Fig.~\ref{fig:sink_detail}.  As described above, multiple
protostars are formed by the initial disk fragmentation ($\Delta t
\lesssim 5,000\,\mr{yr}$), grow in mass via gas accretion ($\Delta t
\sim \mr{a\ few}\times 10,000\,\mr{yr}$), and finally cease to grow due
to radiative feedback, with the accretion rates dropping to such a low
level that the total masses are almost fixed at the end of the
simulations ($\Delta t = 80,000$ or $100,000\,\mr{yr}$).  Many
protostars disappear in mergers induced by unstable gravitational
interaction in a-few-body systems, and only two to four stars survive at
the end of the simulations.

We summarize the properties of the final stellar systems in
Table~\ref{tab:fid}. The total masses are $M_\mr{tot}=100-500\,M_\odot$,
with increasing masses toward higher cloud-scale accretion rates
\citep[e.g.,][]{Hirano:2014aa}.  The number of surviving protostars is
$N_\mr{sink}=2$ or $4$.  The masses of the most and the second most
massive stars in the system, $M_1$ and $M_2$, respectively, are all
quite large ($> 30\,M_\odot$) with the most massive one reaching even
$370\,M_\odot$.  The mass ratio of $q_{12}=M_2/M_1$ is around 0.3-0.8,
which means moderate mass hierarchy.  The semi-major axis of their
binary orbits is $a_{12}=2\times10^3-2\times10^4\,\mr{au}$ at the end of
the simulations.  In all cases, the final outcomes are widely orbiting
($>2,000\,\mr{au}$) massive ($> 30\,M_\odot$) multiple stellar systems.

Although only with very small sample size, it is worth comparing the
distribution of our binary properties with the binary statistics in the
literature, which have been used to estimate the rate of binary black
hole mergers.  Our sample of massive binaries, characterized by large
primary masses ($M_1\gtrsim 60\,M_\odot$), wide orbits
($a_{12}\sim1,000-10,000\,\mr{au}$), and moderate mass ratios
($q_{12}\sim 0.3-0.8$), roughly agrees with the binary statistics of
\citet[see their Fig.~9]{Liu:2021}, who considered the effect of orbital
expansion due to the conservation of angular momentum. The statistics of
\cite{Liu:2021} predict significantly fewer close binaries than other
studies that assume simple models without the above expansion effect
\citep[e.g.,][]{Kinugawa:2014aa,Belczynski:2017aa}.

\subsection{Some characteristic phenomena}
\label{sec:characteristic}

Next, we describe characteristic phenomena observed in our simulation
runs.  Some of them are common to all the runs, while others occur only
in a limited run(s), partly explaining the reason for the observed
similarity or diversity of the formed systems.  In what follows, we
describe those processes one by one in each run.

\subsubsection{Low-\texorpdfstring{$\dot{M}$}{Mdot} run}
\label{sec:low}

\begin{figure}
 \centering \includegraphics[width=5cm]{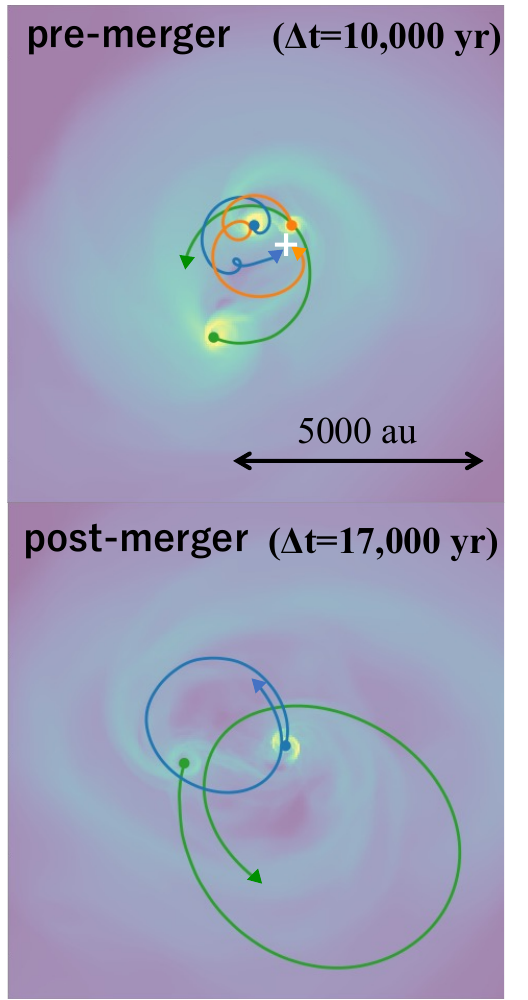}
 \caption{Trajectories of protostars before ($\Delta
 t=10,000\,\mr{yr}$, top) and after ($\Delta t=17,000\,\mr{yr}$, bottom)
 merger in the Low-$\dot{M}$ case. We plot trajectories on the
 background of a white-shaded density map.  Three-body interaction
 induces the merger, leading to the formation of an eccentric
 binary. The white cross in the upper panel indicates the point of
 merger.}  \label{fig:CaseD_merger_traj}
\end{figure}

\begin{figure}
 \centering \includegraphics[width=5.5cm]{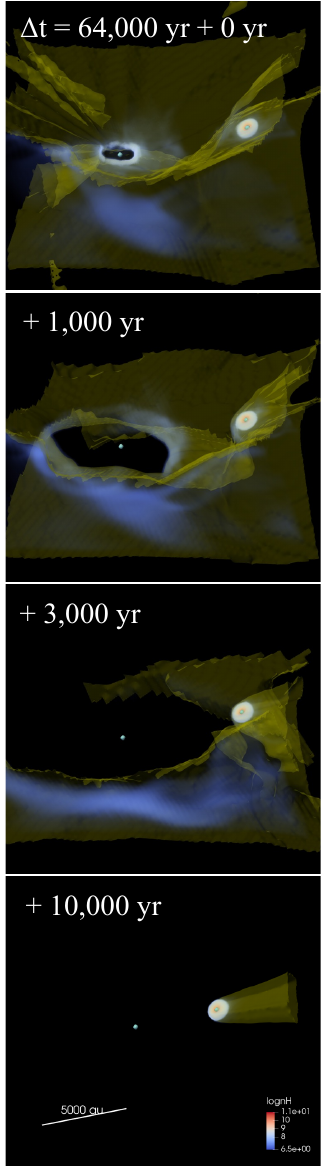}
 \caption{The external photo-evaporation event shown as a time sequence
 in the Low-$\dot{M}$ case. From top to bottom, we show the gas (blue to
 red) and the ionization front (yellow) at the epochs of $0$, $1,000$,
 $3,000$, and $10,000\,\mr{yr}$ after $\Delta t=64,000\,\mr{yr}$, when
 the left protostar begins to completely photo-evaporate its own disk
 from inside.  } \label{fig:external_evap}
\end{figure}

In the Low-$\dot{M}$ run, we identified the following two characteristic
processes, both of which are also observed in the two other cases.  The
first is initial disk fragmentation, which seeds multiple protostars,
and subsequent a-few-body interaction, which induces mergers between
them.  The second is photo-evaporation of a circum-stellar disk by
another protostar.

\paragraph{Initial disk fragmentation followed by a-few-body interaction among fragments}
\label{sec:ini_disk_frag}

At the end of the gravitational collapse, the first protostar appears at
the center of the rotating cloud (see Figs.~\ref{fig:collapse_3d} and
\ref{fig:prof1d_collapse_comparison}).  Around the protostar, an
accretion disk soon forms thanks to the accumulation of the angular
momentum from the rotating envelope.  Since the disk is relatively
massive compared with the central protostar in the early phase
\citep[see][]{Kimura:2021aa}, the disk promptly fragments and produces
two new protostars at $\Delta t \approx 2,000$ and $4,000\,\mr{yr}$
(Fig.~\ref{fig:sink_detail}).  A clump in a spiral arm (at lower-left in
the upper left panel of Fig.~\ref{fig:evolution_faceon_2d} at $\Delta t
= 3,000\,\mr{yr}$) will later develop into the third protostar.

After formation by the disk fragmentation, those three protostars
interact with each other and two of them eventually merge, as seen in
the top panel of Fig.~\ref{fig:CaseD_merger_traj}, which depicts their
trajectories of the before and after the merger.  The trajectories of
the three protostars change violently since a three-body system is
generally unstable.  The interactions tend to be strong in the earliest
phase because the protostars have similar masses and short mutual
distances, comparable to the disk radius. After the merger of two
protostars at $\Delta t \approx 20,000\,\mr{yr}$, a wide-orbit eccentric
binary system is left, as shown in Fig.~\ref{fig:CaseD_merger_traj}
(bottom).

Changing their orbits, the protostars grow in mass by gas accretion, as
seen in Fig.~\ref{fig:sink_detail} (upper left). The accretion rate onto
each protostar remains roughly constant until the merger event, where
the gas in the disks is stripped away by the gravitational interactions
and the accretion rate drops suddenly (Fig.~\ref{fig:sink_detail},
middle left).  Photo-evaporation by the central protostars causes mass
loss from the accretion disk surface (see
Fig.~\ref{fig:evolution_edgeon_2d}, top right) and reduces the accretion
rate even more.  Around this time, the protostars left are scattered
into a low-density peripheral region and gas supply to the disks is also
quenched (see Fig.~\ref{fig:CaseD_merger_traj}, bottom).  Although
accretion is enhanced temporarily due to the perturbation by the other
star on the circum-stellar disk around the pericentric encounter at
$\Delta t \approx 50,000\,\mr{yr}$ \citep{Park:2023ta}, associated mass
growth is modest.

\paragraph{Disk photo-evaporation by a nearby protostar}
\label{sec:ext_pe}

After the pericentric encounter, the accretion rates soon drop again.
One of the stars completes the photo-evaporation of its own accretion
disk and then begins to photo-evaporate the disk of another nearby star.
The time sequence of this ``external'' photo-evaporation is shown in
Fig.~\ref{fig:external_evap}.  Soon after the photo-evaporation of its
own disk by the left star, the ionized region readily expands into a
surrounding lower-density region and finally engulfs the disk on the
right side.

We must remark that the external photo-evaporation in this case does not
play a significant role in setting the final stellar mass.  This is
because, by this time, the accretion rate has already fallen to a value
too low to change the mass.

Note that such external photo-evaporation is observed not only in the
Low-$\dot{M}$ case, but also in all other cases.  In our runs, its role
in controlling the evolution of protostellar systems is limited:
although this may explain peculiar late-time orbital evolution seen in
the Intermediate-$\dot{M}$ case (see Section~\ref{sec:int}), it does not
affect the final stellar mass significantly in any runs.  External
photo-evaporation, however, may play an important role in setting the
mass of low-mass stars that are unable to clear the disk material away
by their own radiation alone.  The reason that such stars are not
observed in our current runs is partly due to resolution limitations.
In any case, given the prevalence of external photo-evaporation, its
role in Pop III star formation is worth further investigation.  For
example, in Galactic star-forming regions, photo-evaporating disks by
external irradiation, also called ``proplyds,'' are of interest because
they not only leave interesting observational tracers but also have
effects on the disk evolution, including planet formation
\citep[e.g.,][]{Yorke:1996aa,Haworth:2017tj}.

\subsubsection{Intermediate-\texorpdfstring{$\dot{M}$}{Mdot} case}
\label{sec:int}

\begin{figure}
 \centering \includegraphics[width=6cm]{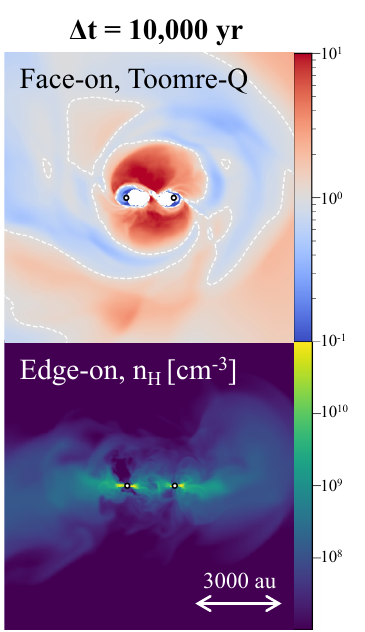}
 \caption{The Toomre's Q parameter (top, face-on view) and number
 density (bottom, edge-on slice) on the scale of the circum-binary disk
 at $\Delta t = 10,000\,\mr{yr}$ for the Intermediate-$\dot{M}$
 case. The dashed contours in the upper panel indicate $Q=1$, below
 which the disk is gravitationally unstable.  The positions of the sink
 particles are indicated with open circles.  } \label{fig:circumbin_2d}
\end{figure}

\begin{figure}
 \centering \includegraphics[width=7cm]{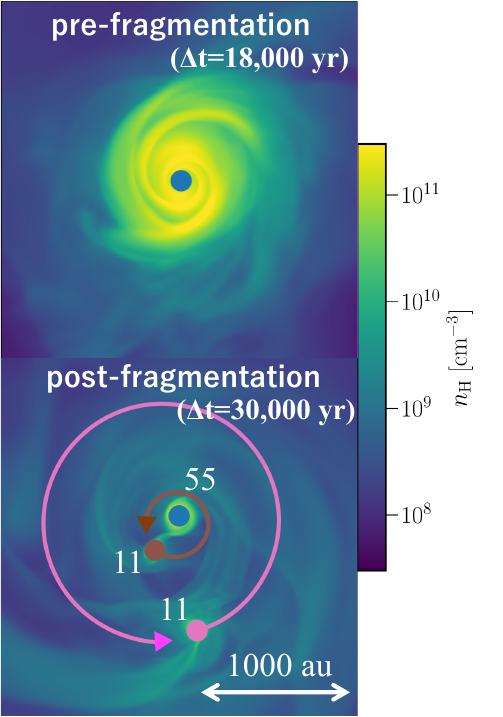}
 \caption{The density distribution of the circum-stellar disk before
 ($\Delta t=18,000\,\mr{yr}$, top) and after ($\Delta
 t=30,000\,\mr{yr}$, bottom) fragmentation, which leads to mini-triplet
 formation in the Intermediate-$\dot{M}$ case.  The gas distribution is
 shown as a face-on slice together with the sink particles marked by
 filled circles.  In the lower panel, the trajectories of the satellites
 are overplotted, in the frame where the barycenter of the inner two
 protostars is fixed, with the masses of the protostars indicated in
 $M_\odot$. } \label{fig:mini3}
\end{figure}

In the Intermediate-$\dot{M}$ case, we also observe the initial disk
fragmentation followed by a-few-body interaction, as well as the
external disk photo-evaporation.  The outcome of the former event is,
however, different here: a circular, rather than eccentric, binary
system is left, unlike in the Low-$\dot{M}$ run, possibly due to the
chaotic nature of a-few-body systems.  The binary then accretes gas from
the circum-binary disk (Figs.~\ref{fig:evolution_faceon_2d}, center) 
around the member stars' circum-stellar disks.  Eventually,
one of them fragments, leading to the formation of a mini-triplet system
(Figs.~\ref{fig:evolution_faceon_2d}, middle right).  Here, we describe
those two processes specific to the Intermediate-$\dot{M}$ case.

\paragraph{Binary evolution by accretion from the circum-binary disk}
\label{sec:bin_acc}

After the initial interaction phase ($\Delta t \gtrsim 6,000\,\mr{yr}$),
the binary's mass and separation significantly change, as shown in
Fig.~\ref{fig:sink_detail}.  This is due to accretion from the
circum-binary disk to the protostars through the circum-stellar disks
(Fig.~\ref{fig:evolution_faceon_2d}, center).  To see this clearly, we
investigate the structure of the accretion flows on the circum-binary
disk scale at $\Delta t = 10,000\,\mr{yr}$.
Fig.~\ref{fig:evolution_faceon_2d} (center) illustrates the binary
system with a separation of $\approx 1000\,\mr{au}$ embedded in a
circum-binary disk spanning from $\approx 2000\,\mr{au}$ to the outer
radius.  This configuration is also evident in the 1D surface density
profile presented in Fig.~\ref{fig:disk1d_cbd} (top) of
Appendix~\ref{sec:cbd}.  Gases are transferred from the circum-binary
disk to the circum-stellar disk of each star through the interfaces,
i.e., on the left (right, respectively) side of the left (right) star,
while gaps are created on the upper and lower sides.

The circum-binary disk rotates at a rate slightly lower than the
Keplerian for the enclosed mass, i.e., the sum of protostars and gas
(see Fig.~\ref{fig:disk1d_cbd} third row in Appendix~\ref{sec:cbd} for
the 1D angular velocity profile).  Since its specific angular momentum
is higher than the binary's orbital value, the accretion from the disk
provides the binary with angular momentum. This explains the observed
increase in binary separation with mass
\citep[e.g.,][]{Munoz:2019aa,Moody:2019aa,Chon:2019aa,Tiede:2020aa,Heath:2020aa}.

Driving mechanism of accretion, i.e., of the angular momentum transfer,
varies depending on the flow segment.  In Fig.~\ref{fig:circumbin_2d}
(top), we present the face-on view of the Toomre's Q parameter, defined
as $Q\equiv c_\mr{s}\,\kappa/\pi\,G\,\Sigma$, with $\Sigma$ the column
density, $\kappa\equiv [(2\Omega/R)
\frac{\mr{d}}{\mr{d}\,R}(R^2\,\Omega)]^{1/2}$ the epicyclic frequency,
and $\Omega$ the angular frequency around the barycenter of the binary.
This parameter is an indicator of gravitational stability: a region of
the disk is stable if $Q>1$ and unstable otherwise.
Fig.~\ref{fig:circumbin_2d} (top) shows that the circum-binary disk is
marginally unstable with $Q\sim 1$ at outer radii $\gtrsim
2000\,\mr{au}$ (also see bottom panel of Fig.~\ref{fig:disk1d_cbd} in
Appendix~\ref{sec:cbd}).  Within the circum-binary disk, the accretion
is driven by the gravitational torque exerted by spiral arms, which are
continuously generated and disrupted, as is commonly assumed in 1D
models of a circum-stellar accretion disk
\citep[e.g.,][]{Matsukoba:2021tb,Kimura:2021aa}.  On the other hand, in
the region between the circum-binary and the circum-stellar disks
($\approx 1500-2000\,\mr{au}$; see Fig.~\ref{fig:evolution_faceon_2d},
center) $Q>1$, i.e., the gas is stable against self-gravity, suggesting
that accretion is primarily driven by the torque exerted by the binary's
gravity.  On the smaller scale within each circum-stellar disk, the
accretion is again governed by disk's self-gravity ($Q\sim 1$), as
presented in Appendix~\ref{sec:csd}.  The circum-stellar disks have a
similar structure to the conventional (quasi-)axisymmetric one in the
literature (e.g., \citealt{Hosokawa:2011aa};
\citetalias{Hosokawa:2016aa}).

The vertical structure of the disks is shown in
Fig.~\ref{fig:circumbin_2d} (bottom).  The circum-binary disk puffs up
to $>1,000\, \mr{au}$ in the vertical direction (also see fourth panel
of Fig.~\ref{fig:disk1d_cbd} in Appendix~\ref{sec:cbd}) due to its weak
gravity, in contrast to the thin circum-stellar disks of a few hundred
au (see Appendix~\ref{sec:csd}).  At this moment, bipolar HII regions
are still confined to the vicinity of the protostars and radiative
feedback is insignificant (Fig.~\ref{fig:evolution_edgeon_2d} center;
see also Appendix~\ref{sec:csd}).

From the analysis above, a unified picture can be drawn regarding
accretion flows, which cause the binary's mass growth, as well as orbit
expansion.  Accretion proceeds first from the circum-binary disk to the
circum-stellar disks and then to the protostars, and the mechanism of
angular momentum transfer depends on the region of the flow.

Finally, we comment on a possible effect of radiative feedback on the
binary orbital evolution.  The increase of the binary separation can be
ascribed to the accretion of high angular momentum gas during the period
$\Delta t \lesssim 30,000\,\mr{yr}$, where the accretion onto stars is
in fact vigorous.  The separation, however, continues to increase
thereafter despite no significant mass growth.  This requires different
explanation.  One possibility is a backward/forward asymmetry in density
around the star due to radiative feedback.  Indeed, black holes moving
in the ambient gas are known to be accelerated by such asymmetries
\citep{Park:2017aa,Toyouchi:2020uf,Sugimura:2020ab}.  Another
possibility is reduction in the gravitational pull on the binary stars
due to the photo-evaporative loss of gas in the region between the
binary.  The effect of radiative feedback on orbital evolution merits
further investigation.
 
\paragraph{Late-time disk fragmentation and acceleration of photo-evaporation}
\label{sec:late_disk_frag}

During binary accretion, one of the circum-stellar disks fragments to
form a mini-triplet system at $\Delta t = 20,000\,\mr{yr}$ (see
Fig.~\ref{fig:sink_detail}).  The state of the disk before ($\Delta
t=18,000\,\mr{yr}$) and after ($\Delta t=30,000\,\mr{yr}$) fragmentation
is shown in Fig.~\ref{fig:mini3}.  The fragmentation of a spiral arm in
the circum-stellar disk (Fig.~\ref{fig:mini3}, top) provides new
protostars (S6 and S7 in Fig.~\ref{fig:sink_detail}, upper middle),
making a mini-triplet system (Fig.~\ref{fig:mini3}, bottom).  The
central protostar is far more massive than the companions, as in
planetary systems.  The mini-triplet inherits the small angular momentum
of the circum-stellar disk and thus has a compact size of $\lesssim
1,000\,\mr{au}$, in contrast to the wide outer binary at a distance of
$6,000\,\mr{au}$, which carries the large angular momentum of the
circum-binary disk.

Here, one circum-stellar disk fragments and the other does not.
Although they are similar before fragmentation ($\Delta
t=18,000\,\mr{yr}$), the former is slightly more unstable with a few
$\times 10\,\%$ more mass than the other and thus causes fragmentation.

Following the formation of the mini-triplet by disk fragmentation, the
gas is quickly lost in the forming stars and their subsequent accretion,
which accelerates the photo-evaporation of the disk (see
Fig.~\ref{fig:mini3}, bottom).  The photo-evaporation of the gas around
the mini-triplet is completed much earlier than the counterpart in the
wide outer binary, at $\Delta t = 50,000\,\mr{yr}$.  After that, the
hierarchical triplet system remains stable and shows no obvious
evolution in our simulation, while the ionized region around the
mini-triplet expands, reaches the other circum-stellar disk, and
photo-evaporates it from the outside, as described in
Section~\ref{sec:low}.

\subsubsection{High-\texorpdfstring{$\dot{M}$}{Mdot} case}
\label{sec:high}

\begin{figure}
 \centering \includegraphics[width=8cm]{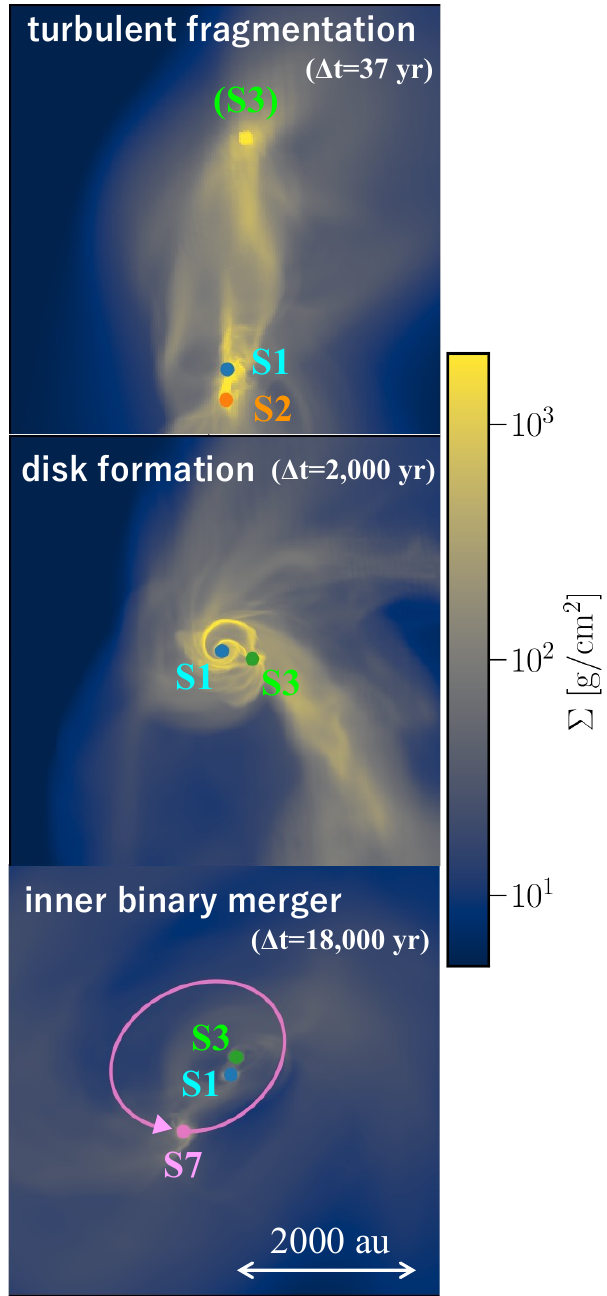}
\caption{The time sequence of initial turbulent fragmentation ($\Delta
t=37\,\mr{yr}$, top), subsequent disk formation ($\Delta
t=2,000\,\mr{yr}$, bottom), and just before the merger of the inner
binary induced by a star in the outer orbit ($\Delta t=18,000\,\mr{yr}$)
in the High-$\dot{M}$ case.  The column density distribution is shown in
the face-on view.  We also show the projected positions of the sink
particles with their ID's.  At the epoch of the top panel, the fragment
labeled with ``(S3)'' is yet to be converted into a sink particle.  The
secondary protostar, S2, in the first epoch merges with the primary, S1,
and disappears before the second epoch. In the bottom panel, we show the
trajectories of protostars in the frame where their barycenter is fixed.
The inner binary later merges due to angular momentum extraction by the
outer protostar.  The axis perpendicular to each image plane is fixed to
the same direction.  } \label{fig:fil_frag}
\end{figure}

In terms of event sequence, the High-$\dot{M}$ case resembles more with
the Low-$\dot{M}$ case than the Intermediate-$\dot{M}$ case.  We observe
initial disk fragmentation and external photoevaporation
(Section~\ref{sec:low}) but do not either accretion from a circum-binary
disk or late-time disk fragmentation (Section~\ref{sec:int}).  As a
process unique to the High-$\dot{M}$ case, below we explain turbulent
fragmentation preceding initial disk formation.

\paragraph{Turbulent fragmentation preceding disk formation}
\label{sec:fil_frag}

We present the gas distribution around protostars in
Fig.~\ref{fig:fil_frag} at the following three stages: initial turbulent
fragmentation ($\Delta t =37\,\mr{yr}$), the subsequent disk formation
($\Delta t= 2,000\,\mr{yr}$), and just before the late-time merger of
the inner binary ($\Delta t= 18,000\,\mr{yr}$).

In the top panel, we can identify a fragment labeled with ``(S3)'' (not
yet converted to a sink particle) on the upper side along a vertical
filamentary structure, in addition to two protostars (S1 and S2) on the
other side. Note that S2 soon merges with S1 at $\Delta t =
100\,\mr{yr}$ and does not appear in Fig.~\ref{fig:sink_detail}, which
starts at $\Delta t = 1,000\,\mr{yr}$.  The turbulent fragmentation
occurs only in this run, presumably because the cloud has strong
turbulence and weak rotation in the initial collapse phase
(Figs.~\ref{fig:collapse_3d} and \ref{fig:prof1d_collapse_comparison}).

Eventually, the accumulation of angular momentum leads to the formation
of a disk around the central protostar (S1), as shown in
Fig.~\ref{fig:fil_frag} (middle).  The disk subsequently becomes
gravitationally unstable and undergoes fragmentation, seeding several
protostars (see Fig.~\ref{fig:sink_detail}, right).  During the disk
formation process, the gas originally belonging to the filament also
accretes onto the disk.  Simultaneously, the fragment, which transforms
into the third sink particle (S3), is gravitationally captured by the
central protostar. 
It then begins orbiting within the disk plane
through the interactions with the disk gas, forming a close binary with
$\sim 300\,\mr{au}$ separation.  The short separation can be attributed
to the relatively low initial angular momentum of the protostar formed
through filament fragmentation, compared with those formed through disk
fragmentation.

After the disk fragmentation, many protostars merge through
gravitational interaction, leaving a three-body system wherein an outer
protostar (S7) orbits an inner binary (S1-S3), as shown in
Fig.~\ref{fig:fil_frag} (bottom).  Soon after the epoch shown in the
figure, however, the long-lasting inner binary, originated from the
capture of a filament fragment around $\Delta t \approx 2,000\,\mr{yr}$,
eventually merges due to the perturbation by S7 at $\Delta t =
20,000\,\mr{yr}$.  This demonstrates that in a hierarchical three-body
system, the outer-orbit star can extract the angular momentum from the
inner binary via gravitational torque and decrease its separation,
rather than increasing the binary separation, as observed in the case of
circum-binary disk accretion (Section~\ref{sec:int}).  We here assumed
that a binary has merged when the separation reaches $128\,\mr{au}$ in
our criterion (see Section~\ref{sec:SFUAMTO}).  In reality, however, our
merger events could represent close-binary formation below the
resolution limit.  This suggests potential importance of turbulent
fragmentation in close-binary formation.

After the merger of the inner binary, an eccentric binary of the merger
product and the outer star is left (Fig.~\ref{fig:evolution_faceon_2d},
lower right), similar to the Low-$\dot{M}$ case.  We observe periodic
modulation of the accretion rates onto the two stars with pericenter
encounters of the binary in Fig.~\ref{fig:sink_detail}
\citep[][]{Park:2023ta}.  While the accretion rates modulate at the same
frequency, their amplitudes are different due to the difference in the
stellar gravity.  The lower-mass star has the accretion rate peaking as
high as $\sim 10^{-2}\,M_\odot/\mr{yr}$, which results in temporal
stellar swelling and significant suppression of UV radiation
(\citetalias{Hosokawa:2016aa}; \citealt{Park:2023ta}; see
Appendix~\ref{sec:proto_stellar_model}).  Later, with gradual decline of
the peak, as well as average, accretion rate, the UV suppression ceases.

\section{Relation between the properties of the final stellar system and natal cloud}
\label{sec:relation}

In all the three runs, we have observed the formation of massive and
wide multiple-star systems, with quantitative variations among them.  In
Section~\ref{sec:cloud_vs_star}, we see the relation between properties
of the final stellar systems, i.e., the total mass and binary
separation, and those of initial clouds.  We then discuss why Pop III
stars predominantly form as massive and wide multiple-stellar systems
based on those relations in Section~\ref{sec:reason}.

\subsection{Dependence of the total mass and binary separation on the cloud properties}
\label{sec:cloud_vs_star}

\begin{figure}
 \centering \hspace*{-0.5cm}
\includegraphics[width=9cm]{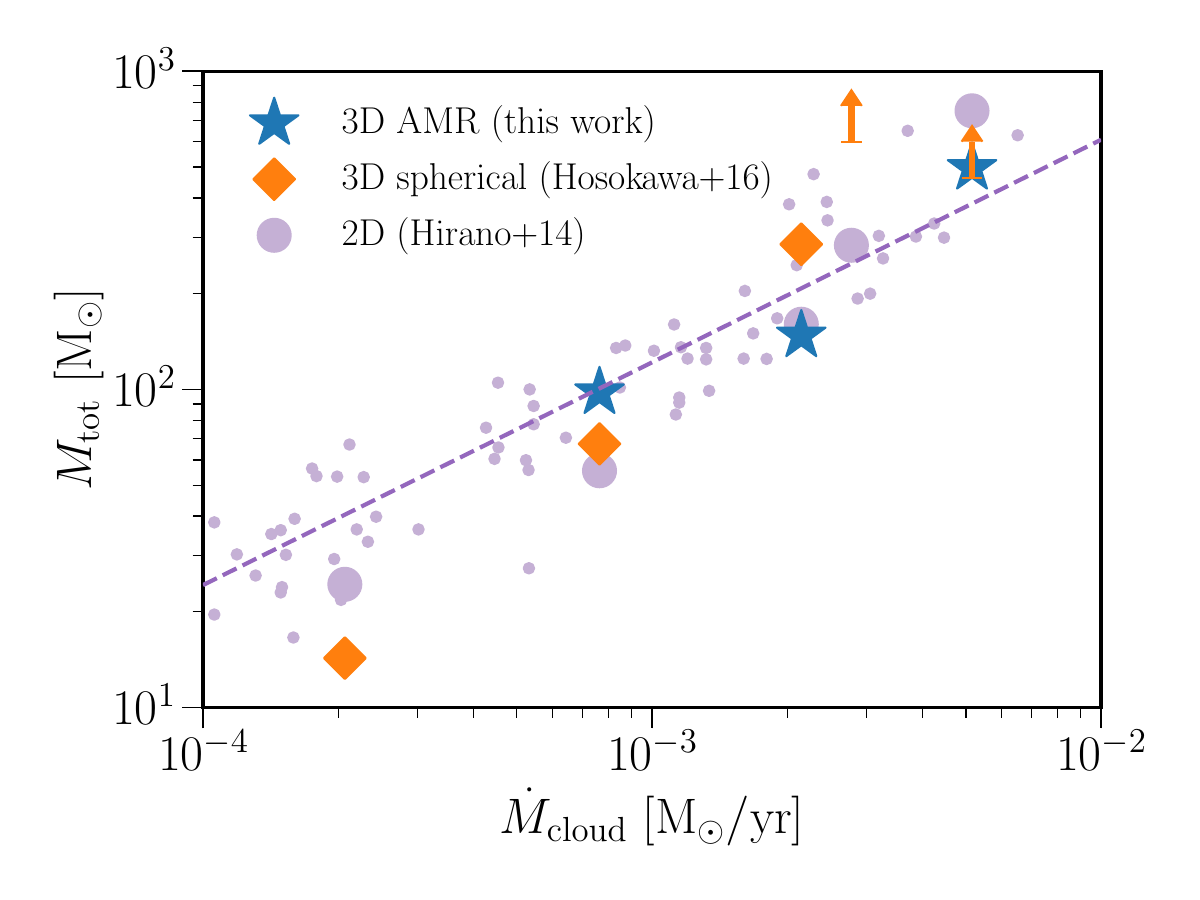}
\caption{The relation between the final total stellar mass, $M_\mr{tot}$
and the initial cloud-scale accretion rate, $\dot{M}_\mr{cloud}$.  We
plot the data from our 3D AMR simulations (blue), along with previous 3D
spherical-grid simulations (\citetalias{Hosokawa:2016aa}, orange), and
previous 2D simulations \citep[][purple]{Hirano:2014aa}.  The 2D data
points are denoted by small symbols, except for those re-examined by 3D
spherical-grid simulations.  The dashed line shows the relation proposed
by \cite{Hirano:2015aa} based on the 2D results (Eq.~\ref{eq:12}).  }
\label{fig:mdot_mass_relation}
\end{figure}

\begin{figure}
 \centering \hspace*{-0.5cm}
\includegraphics[width=9cm]{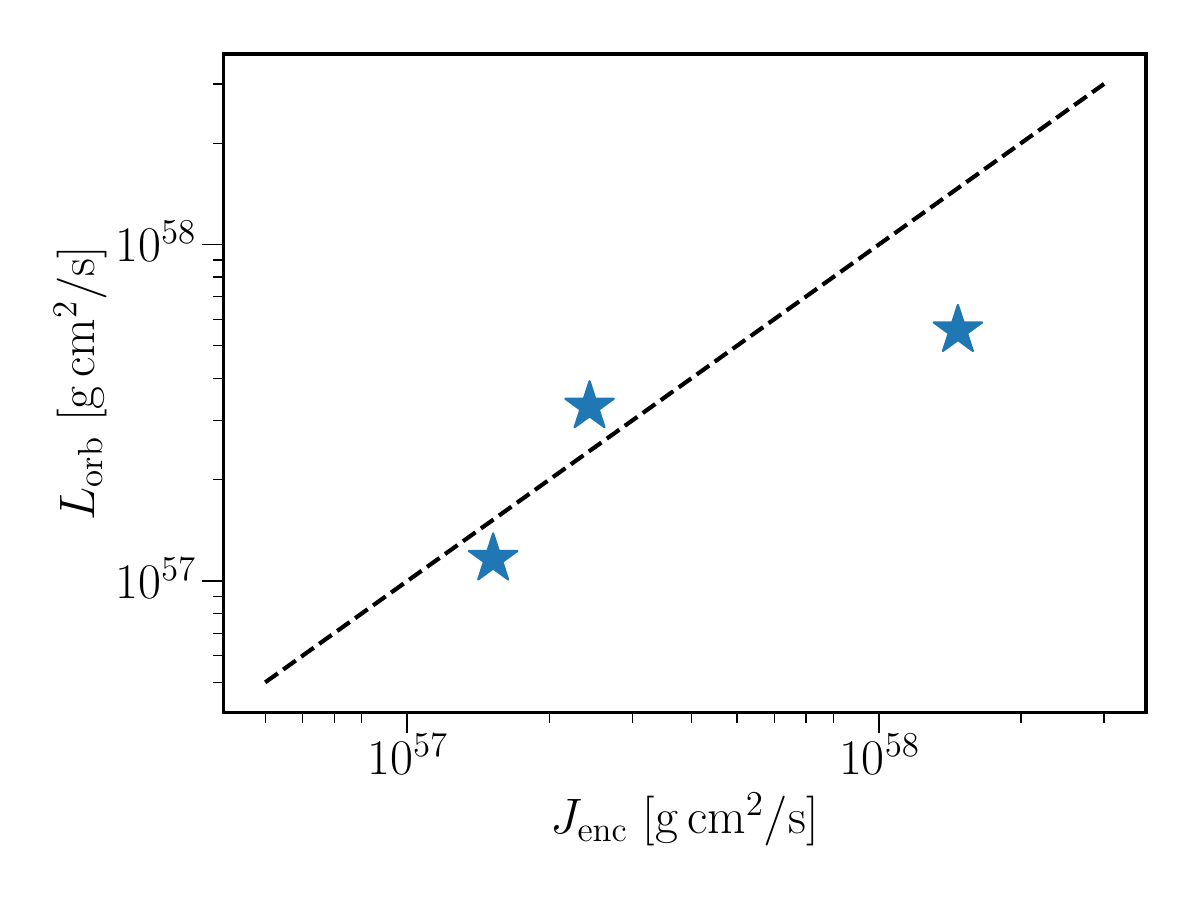}
\caption{The relation between the angular momentum of the binary orbits,
$L_\mr{orb}$, and the corresponding initial enclosed angular momentum of
gas clouds, $J_\mr{enc}$.  The binary's orbital angular momentum and
total mass are evaluated at $\Delta t = 30,000\,\mr{yr}$, before stellar
radiative feedback drives orbital evolution.  The initial enclosed
angular momentum $J_\mr{enc}$ is calculated for the radius $R_\mr{enc}$,
whose enclosed mass is equal to the total mass $M_\mr{tot}$, i.e.,
$J_\mr{enc} \equiv J_{<R_\mr{enc}}$ with $R_\mr{enc}$ satisfying
$M_{<R_\mr{enc}} = M_\mr{tot}$. The plotted points correspond to the
Low-$\dot{M}$, Intermediate-$\dot{M}$, and High-$\dot{M}$ cases from
left to right.  } \label{fig:Jenc_Lorb_relation}
\end{figure}

The relation between the final stellar mass and the initial cloud-scale
accretion rate has been proposed based on 2D simulations of single Pop
III star formation (\citealt{Hirano:2014aa}). Our 3D simulations,
however, have shown that multiple, rather than single, stellar systems
are formed in reality.  Below, we see whether similar correlation still
exists in our case.

Fig.~\ref{fig:mdot_mass_relation} shows the relation between the final
total stellar mass, $M_\mr{tot}$, and the initial accretion rate at the
cloud scale, $\dot{M}_\mr{cloud}$ (see Table~\ref{tab:cloud}).  We plot
our results together with those from previous 2D axisymmetric
simulations \citep{Hirano:2014aa} and 3D spherical-grid simulations
(\citetalias{Hosokawa:2016aa}).  In both cases, only a single star forms
in each cloud due to the numerical limitation, while accretion
modulation induced by disk fragmentation is observed in the latter.  We
find that the total mass tends to be higher for a higher accretion rate
also in our simulations, in agreement well with the fitting function
proposed by \citealt{Hirano:2015aa} (their Eq.7):
\begin{align}
 M_\mr{tot} = 250\,M_\odot \times \left(\frac{\dot{M}_\mr{cloud}}{2.8\times10^{-3}\,M_\odot/\mr{yr}}\right)^{0.7}\,.
\label{eq:12}
\end{align}

Formation of multiple protostars have two effects on the total mass.  On
the one hand, mass sharing among multiple stars leads to smaller mass of
each star and thus weaker radiative feedback (see
Appendix~\ref{sec:proto_stellar_model}).  On the other hand, the
displacement of protostars towards less dense off-centered regions leads
to the suppression of accretion. In addition, the chaotic behavior of
a-few-body systems may also introduce extra scatters in each case
\citep[e.g.,][]{Susa:2019wj,Wollenberg:2020wo}.

The stellar orbits also correlate with the properties of the natal
clouds.  Fig.~\ref{fig:Jenc_Lorb_relation} shows the orbital angular
momentum of the binaries, $L_\mr{orb}$, against the corresponding
initial enclosed gas angular momentum, $J_\mr{enc}$. To determine these
values, we first evaluate the binary's orbital angular momentum and
total mass at $\Delta t = 30,000\,\mr{yr}$.  We then measure the initial
enclosed angular momentum within the radius containing the total mass
(3,000, 4,000, and 9,000$\,\mr{au}$ for the Low-$\dot{M}$,
Intermediate-$\dot{M}$, and High-$\dot{M}$ cases, respectively) for the
cloud profile just before the formation of the first protostar (see
Fig.~\ref{fig:prof1d_collapse_comparison}).  Note that these radii are
somewhat smaller than the cloud size, in which the cloud spin parameter
$\lambda_\mr{cloud}$ is measured in Table~\ref{tab:cloud}.  The
mini-triplet in the Intermediate-$\dot{M}$ case is treated as a single
object put at their barycenter.  This does cause little error in
evaluating the angular momentum of the system as the angular momenta of
the mini-triplet's internal orbits are negligible compared with that of
the wider binary consisting of the virtual body and the other protostar.

We chose the epoch of $\Delta t = 30,000\,\mr{yr}$ in order to eliminate
the influence of radiative feedback on the binary's orbit and focus on
quantities primarily determined by the conservation laws for mass and
angular momentum.  By this epoch, the accretion process is nearly
completed, as shown in Fig.~\ref{fig:sink_detail}.  As we see in the
Intermediate-$\dot{M}$ case, radiative feedback possibly makes the
binary's separation change even after the end of accretion.  Although
some more study on radiative feedback effect on the stellar orbits is
needed, we speculate that its impact on the (already large) binary
separation is rather limited.

Fig.~\ref{fig:Jenc_Lorb_relation} shows that the late-time orbital
angular momentum of the binary/multiple system, roughly agrees with the
initial angular momentum of the cloud.  This means that large angular
momentum/separation of the binaries can be attributed to large angular
momentum of the clouds.  Slightly lower orbital angular momentum in the
High-$\dot{M}$ case than is expected is probably due to strong
turbulence in the initial cloud
(Fig.~\ref{fig:prof1d_collapse_comparison}), which transfers the angular
momentum outward.

\subsection{Reason for massive and wide multiple-star systems}
\label{sec:reason}

When the first protostar is still in the infancy, a disk forms around
it.  Being relatively massive compared with the central star, the disk
easily fragments and forms multiple protostars
\citep[see][]{Kimura:2021aa}.  Such small-body systems are generally
unstable by the gravitational interaction and most of them merge or are
scattered away.  Eventually, only a few protostars can stay within dense
regions near the center and continue to grow in mass.  In the following,
we limit the case of binaries for simplicity although we expect similar
conclusion in a-few body cases.

While the binary system gains mass by accretion from the circum-binary
disk, its mass ratio approaches unity in accordance with previous
dedicated studies
\citep{Bate:1997ab,Satsuka:2017aa,Matsumoto:2019wj,Chon:2019aa}.  Note
that accretion does not cause the merger of pre-existing binaries:
instead it increases the binary separation (see
Fig.~\ref{fig:circumbin_2d} bottom), rather than decreasing it
\citep{Tiede:2020aa,Heath:2020aa}.  Consequently, Pop III stars tend to
form as a system in which a few stars dominate the total mass, and
presumably the total orbital angular momentum.

This system then becomes massive and wide.  The total mass reaches as
high as $80-500\,M_\odot$, thanks to the high cloud-scale accretion rate
(Fig.~\ref{fig:mdot_mass_relation}), which stems from the high
temperature in the primordial gas due to the inefficient cooling
\citep[e.g.,][]{Stahler:1986aa, Omukai:1998aa}.  Simultaneously, the
(outer) binary has a wide orbit with its separation reaching at least
$2,000\,\mr{au}$ due to the large initial angular momentum of the natal
cloud (Fig.~\ref{fig:Jenc_Lorb_relation}) without effective angular
momentum extraction by such processes as magnetic braking or
magnetically-driven outflows
\citep[e.g.,][]{Machida:2008ab,Sadanari:2021aa,Sadanari:2023vd}.  In
this way, Pop III stars predominantly form as multiple systems
consisting of massive stars with wide orbits.

Note, however, that this does not exclude the formation of close
binaries, some of which could be progenitors of binary BH mergers
observed by gravitational waves.  Rather, our results have some
implications for close binary formation.  As in the
Intermediate-$\dot{M}$ run, Pop III stellar systems can be hierarchical,
in which outer wide orbit stars and mini-multiplet systems coexist.
Whereas the separation of the outer binary tends to increase by
accretion from the circum-binary disk, stars in the outer orbit can
extract the angular momentum from the inner binary and shrink its
separation of the inner one, as observed in the High-$\dot{M}$ case.
Although the inner binary was regarded as merged in our simulation, the
actual end product could be a close binary system below our resolution
limit.  Higher resolution simulation is needed in the future to answer
whether such close binaries as gravitational event progenitors are
formed or not among Pop III stars \citep[e.g.][]{Kirihara:2023}.

Note also that low-mass stars may still form although we have not found
them in our simulations, possibly due to our limited spatial resolution
and simplistic sink merger criteria (see also
Sec.~\ref{sec:resolution}). Previous numerical studies on cosmological
Pop III star formation have suggested that numerous stars remains
low-mass ($< 1\,M_\odot$) after ejection from the central region of the
star-forming cloud \citep{Greif:2012aa,Latif:2022wf}. Whether or not
such stars indeed coexist in the system, however, is unlikely to
significantly affect the evolution of massive protostars.  Nevertheless,
their formation and survival are of substantial interest from
observational perspectives as current observations have already ruled
out the cases where an excessive number of low-mass stars survive up to
the present \citep{Hartwig:2015ac, Ishiyama:2016aa}.

In summary, we argue that Pop III star systems are likely comprised of
widely orbiting multiple massive stars. At the same time, these systems
may include embedded close binaries, as well as numerous low-mass stars
ejected from the center.

 \section{Discussion} 
 \label{sec:discussion}

\subsection{Effect of radiative feedback}
\label{sec:rad_feeedback}

\begin{figure}
 \centering
 \includegraphics[width=8cm]{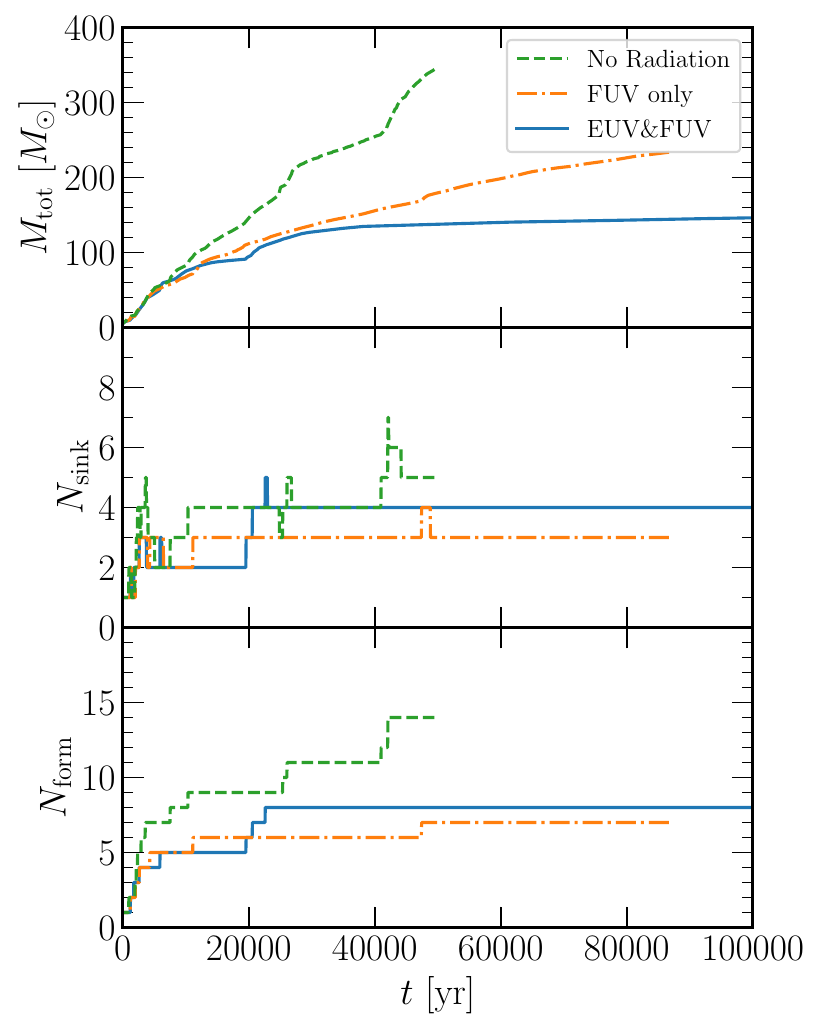}
 \caption{Dependence of protostellar evolution on the prescription of
 radiative feedback. From top to bottom, we plot the total mass
 $M_\mr{tot}$, the number of surviving sinks $N_\mr{sink}$, and the
 number of sinks formed $N_\mr{form}$.  The line types indicate the
 runs: the fiducial run with both EUV and FUV (blue solid), the FUV-only
 run (orange dotted-dash), and the run without radiation (green
 dashed).}  \label{fig:sp_raddep_comparison}
\end{figure}

\begin{figure*}
 \centering
 \includegraphics[width=17cm]{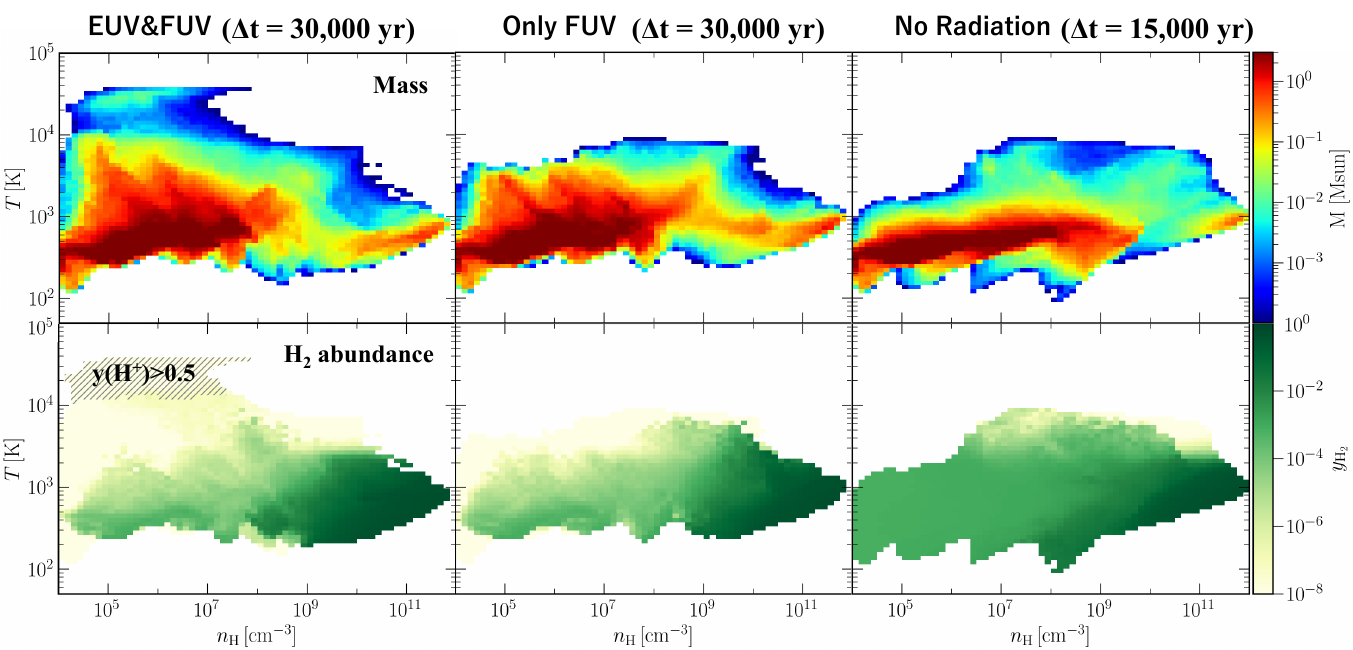}
 \caption{The density-temperature phase diagram in the fiducial run with
 both EUV and FUV (left), the FUV-only run (middle), and the run without
 radiation (right).  We take the data when $M_\mathrm{tot}\approx
 130\,M_\odot$ in all the runs ($\Delta t = 30,000\,\mr{yr}$ in the
 first two runs and $\Delta t = 15,000\,\mr{yr}$ in the last run).  The
 color indicates the mass in the top row and the $\mr{H}_2$ fraction in
 the bottom row.  In the bottom row, hatched is the region where the gas
 is ionized ($y(\mr{H}^+)>0.5$) } \label{fig:nT_FBdep}
\end{figure*}

\begin{table*}
 \centering
 \caption{Summary of simulation setup for additional runs.}
 \label{tab:model_test}
 \begin{tabular}{cccccc} \hline\hline
Run & 
$\Delta x_\mr{min}\, [\mathrm{au}]$ & 
$r_\mathrm{sink}\, [\mathrm{au}]$ & 
$n_\mathrm{sink}\, [\mathrm{cm^{-3}}]$ & 
FUV &
EUV\\
\hline
Int-noRad$^\mr{a}$       & $4$ & $64$ & $1\times10^{11}$ & No & No \\ 
Int-noEUV$^\mr{b}$         & $4$ & $64$ & $1\times10^{11}$ & Yes & No \\ \hline
Int-32au$^\mr{c}$          & $2$ & $32$ & $4\times10^{11}$ & Yes & Yes \\
Int-noRad-32au$^\mr{a,c}$  & $2$ & $32$ & $4\times10^{11}$ & No & No \\  
Int-noRad-16au$^\mr{a,c}$  & $1$ & $16$ & $1.6\times10^{12}$ & No & No \\  
\hline

 \end{tabular}\\
 \begin{flushleft}
Notes. \\
$^\mr{a}$ -noRad denotes no radiation (neither EUV nor FUV).\\
$^\mr{b}$ -noEUV denotes no EUV (FUV only).\\
$^\mr{c}$ 32au and 16au indicate the sink radius.\\
 \end{flushleft}
\end{table*}

The radiative feedback from protostars suppresses the accretion growth
of protostars, as observed in our simulations.  To see how this works
more specifically, we here perform a series of additional runs for the
Intermediate-$\dot{M}$ case with different numerical setups.  Previous
authors suggest that the accretion is suppressed either by FUV
\citep[e.g.,][]{Susa:2013aa,Susa:2014aa} or EUV radiation
\citep[e.g.,][]{McKee:2008aa,Hosokawa:2011aa}.  To discriminate the
effect of each UV component, we perform runs with only FUV (no EUV) and
without radiation (neither EUV nor FUV), in addition to the fiducial run
both with EUV and FUV radiation.  Table~\ref{tab:model_test} summarizes
the setups for the additional runs, along with another series of runs
presented in Section~\ref{sec:resolution}.

Fig.~\ref{fig:sp_raddep_comparison} gives the comparison of the
protostar evolution in the fiducial (EUV and FUV), FUV-only, and
no-radiation runs.
The total mass evolution (top panel) shows variation among the three
runs.  In the no-radiation run, the mass increases linearly at a nearly
constant accretion rate without any feedback.  On the other hand, in the
runs with FUV radiation, the mass growth slows down due to
photodissociation feedback around $\Delta t \sim 10,000\,\mr{yr}$.  In
the FUV-only run, the mass continues to grow at a reduced rate, while
the EUV photoionization feedback nearly halts the accretion at $\Delta t
\sim 40,000\,\mr{yr}$ in the fiducial run with EUV.
The bottom two panels in Fig.~\ref{fig:sp_raddep_comparison} show the
number of surviving sink particles, $N_\mr{sink}$, and the total number
of sinks formed, $N_\mr{form}$.  Due to a-few-body interaction, which
reduces the number of protostars through mergers, $N_\mr{sink}$ is
similar in the range of 3 - 5 in the three runs.
On the other hand, a larger difference is observed in $N_\mr{form}$.
While $N_\mr{form}$ continues to increase with time in the no radiation
run, it reaches a plateau around $\Delta t \sim 20,000\,\mr{yr}$ in both
the fiducial and FUV-only runs.  This plateau is likely caused by the
reduction in the accretion rate onto the disks, which falls below the
rate required to sustain the formation of new sink particles through
disk fragmentation.

To see how radiative feedback affects the thermal state of gases, we
present density-temperature phase diagrams in
Fig.~\ref{fig:nT_FBdep}. The color in each bin represents either the
mass (top) or the $\mr{H}_2$ fraction (bottom).  All the plots are for
the epoch when the total mass is $130\,M_\odot$.  Around this time, FUV
radiation, if included, has already suppressed the accretion and EUV
radiation, again if included, begins to terminate the accretion (see
Fig.~\ref{fig:sp_raddep_comparison}, top).  The corresponding epochs are
$\Delta t = 30,000\,\mr{yr}$ in the fiducial and FUV-only runs and
$\Delta t = 15,000\,\mr{yr}$ in the no-radiation run.

The presence of FUV feedback affects the envelope gas in the density
range of $10^5-10^9\,\mr{cm^{-3}}$.  A substantial portion of the gas in
the envelope is heated to $T\gtrsim1,000\,\mr{K}$ in the runs with FUV,
while most of the envelope gas remains cold at $T\lesssim1,000\,\mr{K}$
in the no radiation run (Fig.~\ref{fig:nT_FBdep}, top).  The higher
temperatures under FUV feedback can be attributed to the reduced
$\mr{H}_2$ fraction and thus cooling due to FUV photodissociation
(Fig.~\ref{fig:nT_FBdep}, bottom).  The suppression of mass growth at
$\Delta t \gtrsim 10,000\,\mr{yr}$ in the runs with FUV radiation
(Fig.~\ref{fig:sp_raddep_comparison}, top) is presumably caused by high
temperature and thus pressure in the envelope.

EUV feedback leads to the formation of bipolar ionized bubbles around
protostars, as depicted in Fig.~\ref{fig:evolution_edgeon_2d}. These
ionized bubbles are observed as a hot
($20,000\,\mr{K}<T<40,000\,\mr{K}$) and low-density
($10^5\,\mr{cm^{-3}}<n_\mr{H}<10^7\,\mr{cm^{-3}}$) component in
Fig.~\ref{fig:nT_FBdep}.  Of course, this component is observed only in
the fiducial run with EUV radiation and not in the runs without EUV
radiation.  Consistent with previous studies
\citep[e.g.,][]{McKee:2008aa,Hosokawa:2011aa}, the EUV photo-evaporation
of accretion disks appears to be responsible for the observed decline in
accretion at $\Delta t \gtrsim 40,000\,\mr{yr}$ in the fiducial run
(Fig.~\ref{fig:sp_raddep_comparison}).

The analysis above highlights the distinct roles played by FUV and EUV
radiation in halting the accretion.  The FUV feedback starts suppressing
the accretion since an early phase by quenching the main coolant
$\mr{H}_2$, but it alone is not able to completely halt the accretion.
Later on, the EUV feedback becomes active and eventually terminates the
accretion by the disk photo-evaporation, finally fixing the stellar
mass.  Given different roles of EUV/FUV radiation in Pop III star
formation, we caution that ignoring either component may lead to
unrealistic results.

\subsection{Resolution effects}
\label{sec:resolution}

\begin{figure}
 \centering
 \includegraphics[width=8cm]{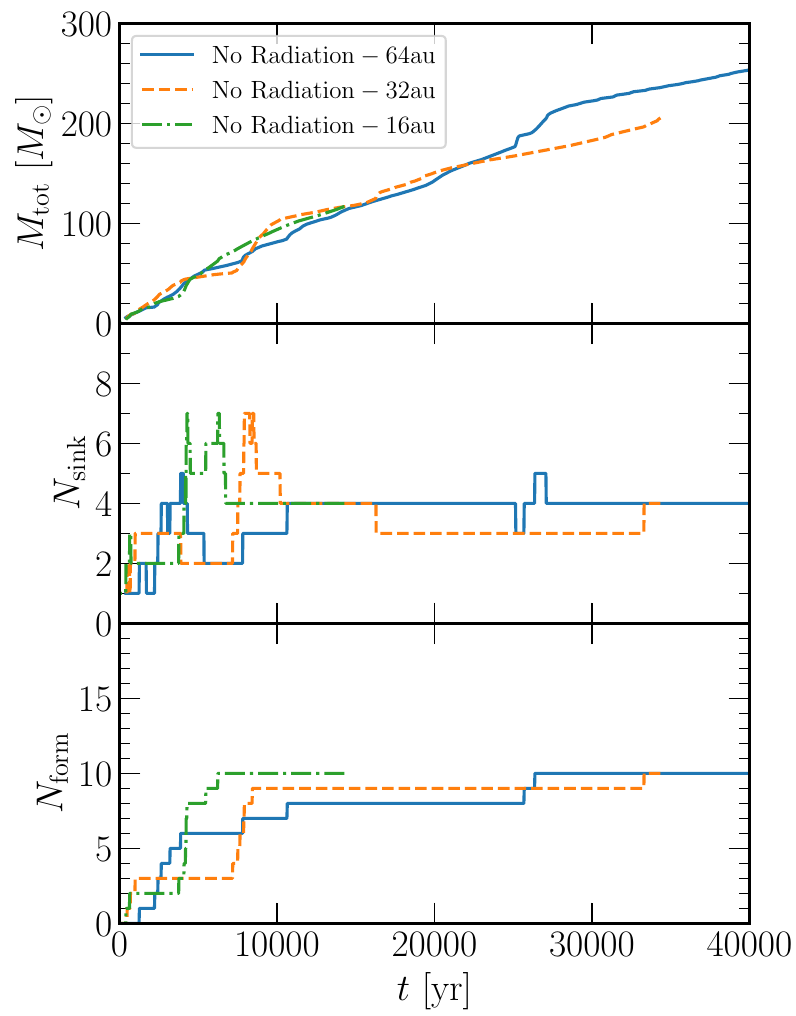}
 \caption{Same as Fig.~\ref{fig:sp_raddep_comparison} but for the
 resolution dependence in no-radiation runs.  We compare the runs
 with different sink radii of $r_\mr{sink} = 64$ (blue), $32$ (orange),
 and $16\,\mr{au}$ (green).  } \label{fig:sp_resdep_comparison}
\end{figure}

\begin{figure}
 \centering
 \includegraphics[width=8cm]{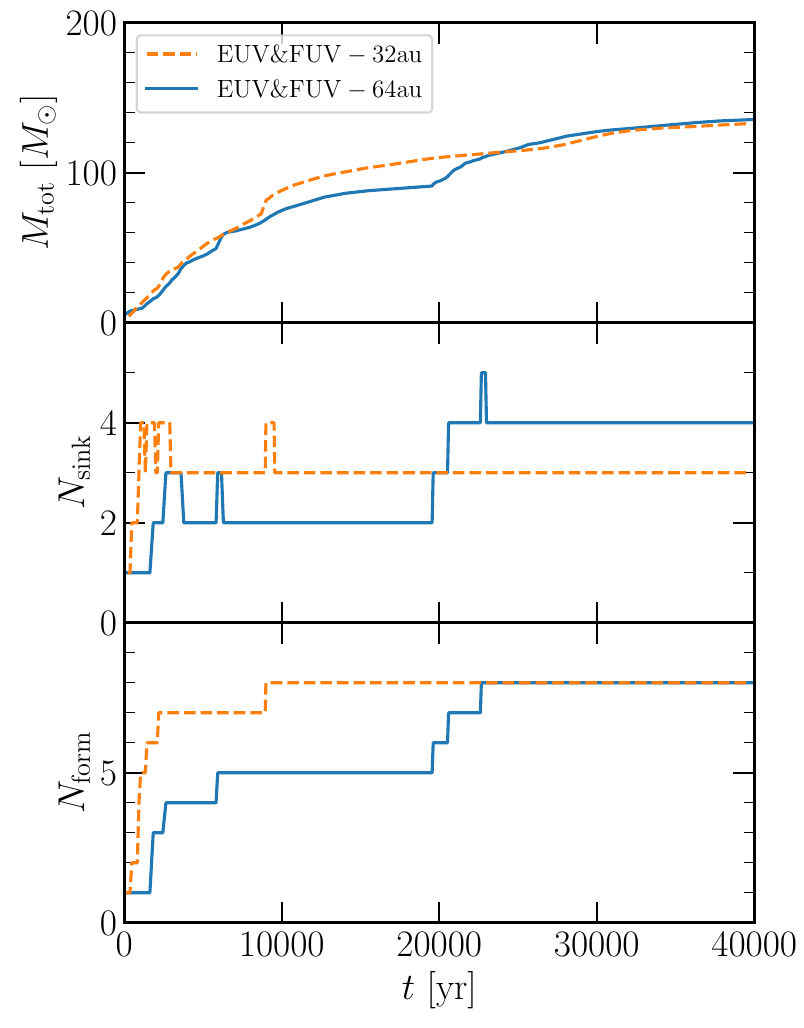}
 \caption{Same as Fig.~\ref{fig:sp_raddep_comparison} but for the
 resolution dependence in full-feedback runs.  We compare the runs with
 different sink radii of $r_\mr{sink} = 64$ (blue) and $32\,\mr{au}$ (orange).  }
 \label{fig:sp_resdep_withrad_comparison}
\end{figure}

To accelerate the computations and follow the accretion evolution to the
end, we introduced sink particles.  Still sink particles at finite
resolution give rise to some uncertainties.  Here, we assess the
resolution effect.

For this purpose, we perform additional runs for the
Intermediate-$\dot{M}$ case with varying resolutions, as summarized in
Table.~\ref{tab:model_test}: two runs without radiation have two and
four times higher resolution and one run with fiducial feedback model,
including both EUV and FUV, has two times higher resolution.  The ratio
of the sink radius to the minimum cell size is fixed at
$r_\mr{sink}/\Delta x_\mr{min}=16$ for all runs, resulting in a sink
radius of $r_\mr{sink}=32\, (16)\,\mr{au}$ in the two (four) times
higher resolution.  Furthermore, we ensure that the Jeans length
(proportional to $(n_\mathrm{sink})^{-2}$ for a constant temperature)
matches $2\,r_\mathrm{sink}$ by increasing $n_\mathrm{sink}$ by a factor
of four (16) when adopting two (four) times higher resolution.

Fig.~\ref{fig:sp_resdep_comparison} shows the resolution dependence of
sink evolution in the runs without radiation, as in
Fig.~\ref{fig:sp_raddep_comparison}.  The total mass, $M_\mr{tot}$, (top
panel) is consistent among runs with varying resolutions.  The number of
surviving sinks (middle panel) is $N_\mr{sink}\sim 4$ due to merger in
all runs while in the bottom panel we observe somewhat earlier rise of
the total number of sinks ever formed, $N_\mr{form}$, for higher
resolution runs.  This can be understood from the fact that, in lower
resolution runs, it takes longer for the disks to acquire enough mass
for fragmentation into sinks.  From the consideration above, we conclude
that our results are insensitive to the resolution within the examined
ranges.

Similarly, we compare the different resolution runs with radiation in
Fig.~\ref{fig:sp_resdep_withrad_comparison}.  Due to limited
computational resources, it is not feasible to continue the high
resolution run beyond $\Delta t =3\times10^4\,\mathrm{yr}$, despite
ongoing accretion (with significant impact of radiative feedback) at
this stage.  Apart from the earlier rise in $N_\mr{form}$ in the higher
resolution run, the evolution of $M_\mr{tot}$, $N_\mr{sink}$, and
$N_\mr{form}$ is in good agreement, as in the cases without radiation.
This also confirms that the resolution dependence in simulations with
radiative feedback is again insignificant within the examined range.

It is nonetheless plausible that some small-scale phenomena near
protostars are not fully captured in our simulations.  For instance,
\cite{Prole:2022uh} reported the formation of more protostars in the
initial disk fragmentation phase at higher resolution than ours.  We,
however, speculate that properties of massive stars would remain largely
unaffected given only a few protostars can grow massive within the dense
central region, as seen in Section~\ref{sec:reason}.  

Note that sink merger criterion is linked to the resolution issue.  We
employ a simplified approach where two sinks are assumed to merge if
their sink spheres overlap.  This criterion grossly simplifies reality,
and the number of surviving sinks should be viewed as a conservative
lower bound.  More stars, and even close binaries below our resolution,
could survive all the way through the end of the accretion phase.

To eliminate all those ambiguities associated with introduction of sink
particles, it is desirable to directly resolve the protostars
themselves. Although still in its infancy, such an attempt is currently
underway \citep[see][]{Kimura:2023}.  Nevertheless, we expect our
conclusion of the formation of massive and wide binary/small-multiple
stellar systems remain unchanged, as discussed in
Section~\ref{sec:reason}.

 \section{Summary and Conclusions} 
 \label{sec:conclusion}

We have studied the formation of Pop III stars by performing radiation
hydrodynamics simulations for three different primordial clouds
extracted from cosmological hydrodynamics simulations.  To accurately
follow the formation of multiple protostars through gas fragmentation
and their radiative feedback on the surrounding gas at a reasonable
computational cost, we have developed a new code SFUMATO-RT, which
employs the adaptive-mesh-refinement and the adaptive-ray-tracing
techniques. Starting from the cloud collapse stage, we follow the growth
of protostars for $\sim 10^5\,\mr{yr}$ until their radiative feedback
quenches the accretion and nearly fixes the stellar properties.  Below,
we summarize our findings.

\renewcommand{\labelenumi}{(\roman{enumi})}
\begin{enumerate}
\item Pop III stars predominantly form as massive and wide
      binary/small-multiple-star systems in all three cases, with masses
      of $30-370\,M_\odot$ and separations of
      $2\times10^3-2\times10^4\,\mr{au}$ (Section~\ref{sec:results}).

\item The properties of the formed stellar systems correlate with those
      of their natal clouds (Section~\ref{sec:cloud_vs_star}): the total
      mass of the system increases with the cloud-scale accretion rate,
      and the angular momentum of the binary orbit matches that of the
      cloud.

\item The total mass of the formed stellar system is consistent with
      that in previous simulations of single-star formation
      (\citealt{Hirano:2014aa,Hirano:2015aa};
      \citetalias{Hosokawa:2016aa}). The individual masses decrease due
      to mass sharing among the multiple stars.

\item The reason for the formation of massive and wide
      binary/small-multiple star systems is discussed in
      Section~\ref{sec:reason}. The large total mass is due to the high
      accretion rate of the primordial gas, while the large angular
      momentum (i.e., large separation) is due to the absence of
      effective angular momentum transfer mechanisms via magnetic
      fields.

\item The following processes are identified as characteristic of Pop
      III star formation:
\end{enumerate}

\begin{itemize}

 \item A disk forms around the first protostar in the simulation. It
       then fragments and seeds multiple protostars.  Subsequent
       unstable a-few-body interactions lead to the merger or scattering
       of most protostars from the central region, limiting the number
       of protostars that grow into massive stars.  Consequently, Pop
       III stars form as a multiple-stellar system, with a few stars
       dominating the system's total mass.

 \item Protostars that complete the photo-evaporation of their own disks
       earlier than the others subsequently externally photo-evaporate
       the disk around another protostar in their vicinity. While the
       impact of this process on the final stellar system requires
       further investigation, it may explain the observed late-time
       orbital evolution in one of the simulations. Moreover, external
       photo-evaporation may influence the evolution of Pop III disk-star
       system, as explored in the context of present-day star formation.

 \item When binary protostars in a circular orbit form, they accrete gas from the circum-binary disk through
       their respective circum-stellar disks, resulting in an increase
       in their mass and separation due to the accretion of high angular
       momentum gas. The fragmentation of a circum-stellar disk leads to
       the formation of a mini-multiple-star system and accelerates the disk
       photo-evaporation. The mini-multiple stars have shorter
       separations than the original wide-orbit binary stars,
       suggesting that the late-time fragmentation of circum-stellar
       disks may be one of the mechanisms for the formation of
       short-orbit systems.

 \item With strong initial turbulence, turbulent fragmentation occurs
       before disk formation. A protostar seeded by turbulent
       fragmentation is captured by and eventually merges with the
       central protostar due to its small angular momentum.  The capture
       of stars originating from turbulent fragmentation may be another
       mechanism for the formation of short-orbit systems.

\end{itemize}

Our findings have two major implications.

First, the reduction in the individual masses of Pop III stars should
change their role in driving the subsequent evolution of the Universe
through stellar radiation, supernovae, and seeding BHs that might
rapidly grow into supermassive BHs \citep[see, e.g., ][and references
therein]{Sugimura:2017ab,Sugimura:2018aa,Sugimura:2020ab}.
Incorporating our findings into cosmological simulations of galaxy
formation is crucial for unveiling the early evolution of the Universe
\citep[see, e.g.,][]{Garcia:2023tg}.

 Second, the binary Pop III stars in our simulations are massive enough
($> 30\,M_\odot$) for the merger of the remnant BHs to be detectable
with current gravitational wave detectors if they merge
\citep[e.g.,][]{Kinugawa:2014aa}.  Their separations are too large
($\sim1,000-10,000\,\mr{au}$) for the remnant BHs to merge solely
through the binary interaction within the age of the present
Universe. However, if these binaries migrate into nuclear stellar
clusters during the cosmic structure formation process, dynamical
hardening can substantially accelerate the merger, as proposed by
\cite{Liu:2021a,Liu:2020a}.  Furthermore, our simulations suggest the
possible presence of mini-binaries embedded in a wider system.  Such
mini-binaries may have separations short enough ($\lesssim 1\,\mr{au}$)
for the remnant BHs to merge via binary interaction
\citep[e.g.,][]{Kinugawa:2014aa,Tanikawa:2021aa}. In any case, further
high-resolution simulations explicitly resolving such close binaries are
necessary to clarify the origin of gravitational wave events.  origin of
gravitational wave events.

While we have consistently treated the radiation feedback from multiple
protostars, we have not considered potentially important effects, such
as a magnetic field and a background radiation field.  While the
strength of the primordial magnetic field is still unknown, it is
proposed that the magnetic field is generated and amplified by
small-scale dynamo during minihalo formation
\citep{Xu:2008tb,Schleicher:2010,Sur:2010aa,Turk:2012aa,Schober:2012tb,McKee:2020ve,Stacy:2022ua}.
If this is the case, we need to consider magnetic field effects in Pop
III formation simulations
\citep[see][]{Machida:2006aa,Sharda:2020vd,Sharda:2021tc,Sadanari:2021aa,Sadanari:2023vd,Prole:2022tq,Stacy:2022ua,Saad:2022vg,Hirano:2022vs}.
Some Pop III stars are thought to form under the influence of radiation
fields from earlier stars. Various types of radiation, including X-rays,
EUV, and FUV, can affect the Pop III star formation
\citep[e.g.,][]{Hummel:2015aa,Park:2021aa,Park:2021ab,Park:2023ta}.

The recent launch of the {\it James Webb Space Telescope} (JWST) has
accelerated the observational quest into the early Universe. To prepare
for the wealth of upcoming observational discoveries, it has become more
urgent than ever to unveil the early cosmic history through simulations,
including Pop III star formation, which represents the very first step
toward the formation of various objects in the Universe.

While our simulations have advanced our understanding of Pop III star
formation, there is a need to increase the sample size to study the
diversity in Pop III star formation.  Furthermore, moving to higher
resolutions is crucial for a more realistic depiction of Pop III star
formation. In our future work, we will extend our research in the directions of higher resolutions and larger sample sizes.

\begin{acknowledgments}
The authors wish to express their cordial gratitude to Prof. Takahiro
Tanaka, PI of Innovative Area Grants-in-Aid for Scientific Research
``Gravitational wave physics and astronomy: Genesis'', for his
continuous interest and encouragement.  The authors also thank Kunihito
Ioka, Jeong-Gyu Kim, Kazutaka Kimura, Masahiro Machida, Takashi Okamoto,
Jongwon Park, Massimo Ricotti, Kenji Eric Sadanari, Masaru Shibata,
Hajime Susa, Kengo Tomida, Masauyuki Umemura, Hidenobu Yajima, and Naoki
Yoshida for fruitful discussions and comments.  This work has never been
accomplished without the support by MEXT/JSPS KAKENHI Grant Number
21K20373 (KS), 17H02863, 17K05394, 18H05436, 18H05437, 23K03464 (TM),
19H01934, 21H00041 (TH), 18J01296, 18H05222, 21H01123, 21K13960 (SH),
and 17H01102, 17H02869, 17H06360 (KO).  The numerical simulations were
performed on the Cray XC50 {\tt Aterui II} at CfCA of the National
Astronomical Observatory of Japan, the computer cluster {\tt Draco} at
Frontier Research Institute for Interdisciplinary Sciences of Tohoku
University, and the Yukawa-21 at Yukawa Institute for Theoretical
Physics at Kyoto University.  This work is also supported by the Hakubi
Project Funding of Kyoto University (KS).
\end{acknowledgments}

  \appendix

  \appendix

  \section{Chemical and thermal modeling} 
  \label{sec:chem_therm_models}

  \subsection{Chemical network} 
  \label{sec:chem_network}

  \begin{table}[t]
   \centering
   \caption{Chemical Reactions}
   \label{tb:chem}
   \begin{tabular}{llc}
    \hline
    \hline
    No. & Reactions & References \\
    \hline
    1 &  H      +  e  $\rightarrow$ H$^+$ + 2 e        &  1 \\
    2$^\mr{a}$ &  H$^+$  +  e  $\rightarrow$ H     + $\gamma$   &  2 \\
    3 &  H$^-$  +  H  $\rightarrow$ H$_2$ + e   &  3 \\
    4 &  H$_2$  +  H$^+$ $\rightarrow$ H${_2}^+$ + H   &  4 \\
    5 &  H$_2$  +  e     $\rightarrow$ 2 H + e         &  4 \\
    6 &  H$_2$  +  H  $\rightarrow$  3 H        &  5 \\
    7 &  3 H          $\rightarrow$  H$_2$ + H  &  6 \\
    8 &  2 H  + H$_2$ $\rightarrow$  2 H$_2$           &  7 \\
    9 &  2 H$_2$      $\rightarrow$  2 H   + H$_2$     &  7 \\
    10 & H    + e     $\rightarrow$  H$^-$ + $\gamma$  &  4 \\  
    11 & H + $\gamma$ $\rightarrow$  H$^+$ + e         &  8 \\
    12$^\mr{b}$ & H$_2$ + $\gamma$ $\rightarrow$ 2 H            &  9 \\ 
    13 & 2 H          $\rightarrow$  H$^+$ + e + H     &  7 \\ 
    14 & H$^-$ + e    $\rightarrow$  H + 2 e           &  1 \\
    15 & H$^-$ + H$^+$ $\rightarrow$ H$_2^+$ + e       & 4 \\
    16 & H$^-$ + H$^+$ $\rightarrow$ 2 H               &  4 \\
    17 & H$^-$ + $\gamma$ $\rightarrow$ H + e          & 10 \\
    18 & H + H$^+$ $\rightarrow$ H${_2}^+$ + $\gamma$  & 4 \\
    19 & H${_2}^+$ + H $\rightarrow$ H$_2$ + H$^+$     & 4 \\
    20 & H${_2}^+$ + e $\rightarrow$ 2 H               & 4 \\
    21 & H${_2}^+$ + H$^-$ $\rightarrow$ H$_2$ + H     & 11 \\
    \hline
   \end{tabular}
   \begin{flushleft}
    NOTES. \\
    $^\mr{a}$ Case B recombination rate. \\

    REFERENCES.\textemdash 
    (1) \cite{Abel:1997aa};
    (2) \cite{Glover:2007aa};
    (3) \cite{Kreckel:2010aa};
    (4) \cite{Galli:1998aa};
    (5) \cite{Martin:1996aa};
    (6) \cite{Forrey:2013aa};
    (7) \cite{Palla:1983vg};
    (8) \cite{Osterbrock:2006aa};
    (9) \cite{Draine:1996aa};
    (10) \cite{John:1988aa};
    (11) \cite{Millar:1991wg}.
   \end{flushleft}
  \end{table}

Our chemical network is summarized in Table \ref{tb:chem}.  It is the
same as in \citetalias{Hosokawa:2016aa}, except different treatment of
$\mathrm{H}_2$ photoionization. The model of $\mathrm{H}_2$
photoionization has an ambiguity since $\mathrm{H}_2$ direct
photoionization requires an energy of at least 15.2 eV but we have only
one frequency bin for EUV photons ($h\nu>13.6\,\mathrm{eV}$).  In
contrast to \citetalias{Hosokawa:2016aa}, we forbid $\mathrm{H}_2$
direct photoionization and allow $\mathrm{H}_2$ to be photoionized by a
two-step process: $\mathrm{H}_2$ photodissociation
$\mathrm{H}_2+\mathrm{FUV} \rightarrow 2\,\mathrm{H}$ followed by H
photoionization $\mathrm{H}+\mathrm{EUV}\rightarrow \mathrm{H}^+ +
\mathrm{e}$.

  \subsection{Thermal processes} 
  \label{sec:thermal_proc}

  \begin{table}
   \centering
   \caption{Heating and Cooling Processes}
   \label{tab:cool_rates}
   \begin{tabular}{llc} \hline
    No. & Process  & Ref. \\ \hline
    \multicolumn{3}{c}{Heating}\\ \hline
    1 & $\mr{H}$ photoionization & 1\\ \hline
    \multicolumn{3}{c}{Cooling}\\ \hline
    2$^{a}$ & $\mr{H}_2$ collisional excitation & 2,3\\
    3$^{b}$ & $\mr{H^{-}}$ free-bound emission & 4\\
    4 & $\mr{H}$ collisional ionization & 5\\
    5 & $\mr{H}$ collisional excitation & 5\\
    6$^{c}$ & Compton scattering & 5 \\
    7 & $\mr{He^{+}}$ collisional excitation & 5\\
    8 & $\mr{H}$ free-free emission & 6\\
    9 & $\mr{H}$ free-bound emission & 7\\ \hline
    \multicolumn{3}{c}{Heating/cooling}\\ \hline
    10$^{a,d}$ & Chemical heating/cooling & 8, 9 \\
    11 & Compression heating/expansion cooling &  \\\hline

   \end{tabular}\\
   \begin{flushleft}
    NOTES.\textemdash 
    $^{a}$ We calculate the line escape probability using a fitting formula provided in \cite{Fukushima:2018aa} with $\mr{H_2}$ column density estimated as $N_\mr{H_2}=\lambda_\mr{J}\,n_\mr{H}\,y(\mr{H_2})$.\\
    $^{b}$ We assume the average photon energy $\langle h\nu\rangle = k_\mathrm{B}T + \chi_\mathrm{H^{-}}$.\\
    $^{c}$ We assume the CMB temperature $T_\mr{CMB}=2.73(1 + z)$ with $z=15$.\\
    $^{d}$ The binding energies are $\chi_{\mr{H}}=13.6\,\mr{eV}$, $\chi_{\mr{H}_2}=4.48\,\mr{eV}$, and $\chi_{\mr{H}^{-}}=0.75\,\mr{eV}$ \citep{Omukai:2001aa}\\
    REFERENCES.\textemdash 
    (1) \cite{Osterbrock:2006aa};
    (2) \cite{Glover:2015ab};
    (3) \cite{Fukushima:2018aa}
    (4) \cite{John:1988aa};
    (5) \cite{Anninos:1997aa};
    (6) \cite{Glover:2007aa};
    (7) \cite{Sugimura:2017ab};
    (8) \cite{Hollenbach:1979aa};
    (9) \cite{Shapiro:1987aa}.
   \end{flushleft}
  \end{table}

  The heating and cooling processes considered in this paper are
  summarized in Table~\ref{tab:cool_rates}. We add to the thermal model
  of \citetalias{Hosokawa:2016aa} the $\mr{H}$ free-free emission and
  the $\mr{H}$ free-bound emission cooling.  Following
  \citetalias{Hosokawa:2016aa}, we do not consider the $\mathrm{H}_2$
  CIE cooling, which is ineffective in the density range of our
  simulations ($n_\mr{H}<10^{12}\,\mr{cm^{-3}}$), and becomes the
  dominant coolant only at higher densities
  \citep[$n_\mr{H}\gtrsim10^{13}\,\mr{cm^{-3}}$; see][]{Omukai:2001aa}.

  \section{Protostar model} 
  \label{sec:proto_stellar_model}

\begin{figure}
   \centering
   \includegraphics[width=7.5cm]{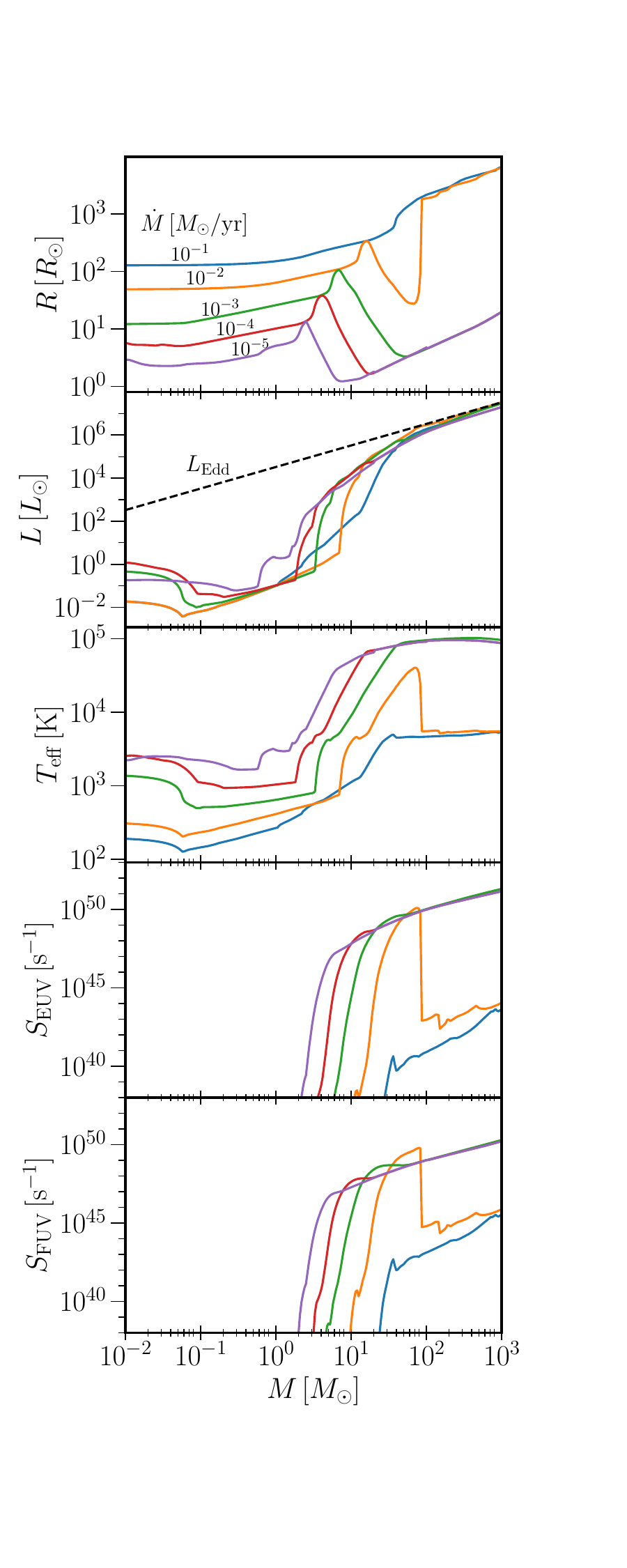}
   \caption{Radiative properties of Pop III protostars obtained from the
   table used in this work.  We plot the protostellar radius $R_*$,
   luminosity $L_*$, effective temperature $T_\mr{eff}$, EUV emissivity
   $S_\mr{EUV}$, and FUV emissivity $S_\mr{FUV}$ from top to bottom.
   The horizontal axis corresponds to the mass $M$, and the color
   indicates the accretion rate
   $\dot{M}=10^{-5}\,,10^{-4}\,,10^{-3}\,,10^{-2}\,,\mr{and}\,10^{-1}\,M_\odot/\mathrm{yr}$
   (see labels in the top panel).  In the second panel, we also show the
   Eddington luminosity $L_\mr{Edd}$ with a dashed line.}
   \label{fig:stellar_model}
  \end{figure}

  We evaluate the radiation from Pop III protostars using the
  pre-calculated table based on one-dimensional stellar evolution
  calculations \citep{Hosokawa:2009aa,Hosokawa:2010aa}.  The table gives
  the protostellar radius $R_*$ and luminosity $L_*$ for a given mass
  $M_*$ and accretion rate $\dot{M}$.  Using $R_*$ and $L_*$, and
  assuming the blackbody spectrum with the effective temperature
  $T_\mathrm{eff} = (L_*/4\pi\,R_*^2\sigma_\mathrm{SB})^{1/4}$ with the
  Stefan–Boltzmann constant $\sigma_\mathrm{SB}$, we calculate the EUV
  emissivity $S_\mathrm{EUV}$ and the FUV emissivity $S_\mathrm{FUV}$ as
  \begin{align}
   S_\mathrm{EUV} &= 4\pi\, R_*^2 \int_{13.6\mathrm{eV}}^\infty \frac{\pi\,B_\nu(T_\mathrm{eff}))}{h\nu}\mathrm{d}\nu\,,\\
   S_\mathrm{FUV} &= 4\pi\, R_*^2 \int_{12.4\mathrm{eV}}^{13.6\mathrm{eV}} \frac{\pi\,B_\nu(T_\mathrm{eff}))}{h\nu}\mathrm{d}\nu\,.
   \label{eq:seuv_sfuv}
  \end{align}
  We present our protostellar model in Fig.~\ref{fig:stellar_model}.

  \section{Adaptive-Ray-Tracing implementation} 
  \label{sec:ART}

  We implement an ART module on SFUMATO, following the previous
  implementations on other codes
  \citep{Abel:2002ab,Krumholz:2007ab,Wise:2011aa,Rosen:2017aa,Kim:2017aa}.
  Here, we briefly review our ART implementation, optimized for
  SFUMATO's oct-tree-type grid structure \citep[see][for
  detail]{Matsumoto:2007aa}. In SFUMATO's terminology, grids are the
  collection of cells (each grid has $16^3=4096$ cells in our
  simulations).  Depending on a given refinement condition, each grid is
  refined to $2^3=8$ self-similar daughter grids.

  In our implementation, each ray structure holds a set of variables,
  including the grid ID, the source ID, the direction specified by the
  HEALPix ID and level \citep[see][]{Gorski:2005aa}, the distance from
  the source, and the optical depth in each frequency. The ray position
  is uniquely determined by the combination of source ID, direction, and
  distance.  Each ray belongs to one of the CPUs.  The CPU stores the
  ray either in the job list or in a list for sending prepared for each
  recipient CPU, depending on which CPU is responsible for the ray's
  grid ID. The ray lists are implemented as linked lists.

  When distributing a ray at a radiation source, we initialize the ray
  variables with its grid ID determined by searching for it from the ray
  position.  We follow the same procedure at ray splitting, which is
  implemented as the annihilation of one parent ray and the generation
  of four daughter rays at one higher HEALPix level.
  
  Each CPU traces rays in its job list by picking up a ray from the list
  one by one and advancing it.  At each ray advancement, the ray passes
  through the cells in the grid specified by the ray's grid ID until it
  reaches a grid boundary, where the ray is assigned a new grid ID
  corresponding to the adjacent grid to which the ray will move.  After
  the advancement, the ray is returned to the job list if the new grid
  ID belongs to the same CPU or is put in a list for sending otherwise.

  The rays in a list for sending are sent to the recipient CPU at some
  intervals.  Currently, we let each CPU send rays when either its job
  list is empty, more than 200 rays are stored in one of the lists for
  sending, or 2000 times ray advancements have been made. The recipient
  CPU receives the rays and stores them in its job list.

  To check the completion of ray tracing, we define a total job size as
  $N_\mr{ray,tot}=12\times2^{2\,l_\mr{ray,max}}$, which is the number of
  rays if all rays are split to the maximum HEALPix level
  $l_\mr{ray,max}$.  In this work, we set $l_\mr{ray,max}=15$, which is
  large enough to ensure that rays are always split as long as the
  ray-splitting condition is satisfied.  We terminate a ray either when
  it reaches a boundary of the computational box or when it is
  sufficiently attenuated (specifically, when $\tau_\mr{EUV}>100$ and
  $N_\mr{H_2}>10^{22}\,\mr{cm^{-2}}$ ($f_\mr{shield}<10^{-8}$) in this
  work).  When a ray at a level $l_\mr{ray}$ is terminated, the
  corresponding job size, $N_\mr{ray}=12\times2^{2\,l_\mr{ray}}$, is
  added to each CPU's job completion counter, $N_\mr{ray,cpu}$.  At some
  intervals, the CPUs check via MPI communication whether the sum of
  $N_\mr{ray,cpu}$ over all CPUs is equal to $N_\mr{ray,tot}$. If this
  is the case, the ART procedure is completed.

  We have performed several tests to check the validity of our ART
  implementation, including tests of static and expanding HII bubbles
  around single radiation sources and a test of static overlapping HII
  bubbles around two radiation sources.  All test results are physically
  reasonable or agree well with analytical estimates, confirming the
  validity of our implementation.

 \section{1D profile of a circum-binary disk}
 \label{sec:cbd}

\begin{figure}
 \centering \includegraphics[width=7cm]{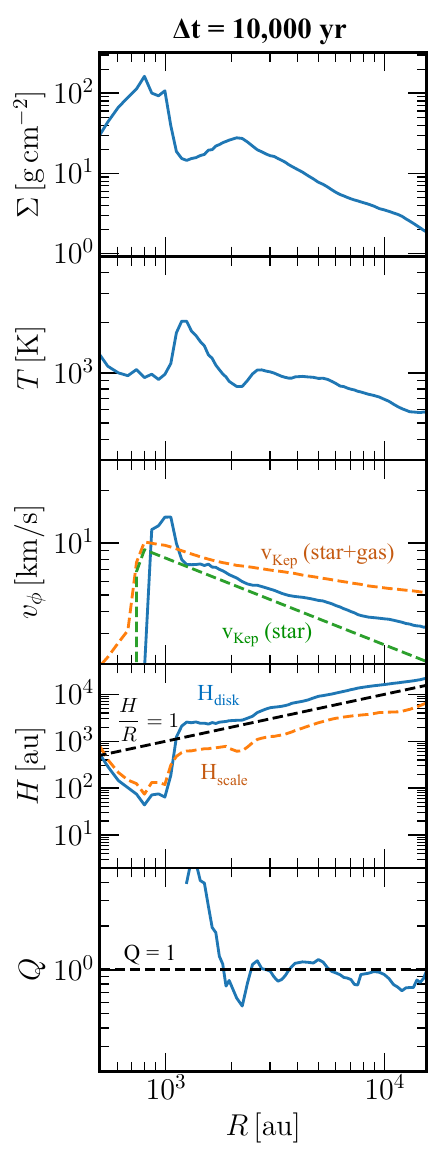}
 \caption{One-dimensional profile of the circum-binary disk presented in
 Fig.~\ref{fig:circumbin_2d}. From top to bottom, we plot the
 angle-averaged column density $\Sigma$, temperature $T$, angular
 velocity $v_\phi$, disk height $H_\mr{disk}$ and scale height
 $H_\mr{scale}$ (see text for definitions), and Toomre's Q parameter
 $Q$.  In the third panel, we show the two Kepler velocities considering
 only the mass of the protostars (green) and considering both the mass
 of the protostars and the enclosed gas mass (orange).  Black dashed
 lines indicate the aspect ratio of unity, $H/R=1$, in the fourth panel
 and the marginally stable state, $Q=1$, in the last panel.  The two
 sink particles are located around 800$\,\mr{au}$ from the center.}
 \label{fig:disk1d_cbd}
\end{figure}

To study more quantitatively the accretion flows on the scale of the
circum-binary disk illustrated in Figs.~\ref{fig:evolution_faceon_2d}
(center), \ref{fig:evolution_edgeon_2d} (center), and
\ref{fig:circumbin_2d}, we present the 1D profile of angle-averaged disk
quantities in Fig.~\ref{fig:disk1d_cbd}. To obtain the disk quantities,
we first set the barycenter as the center and the orbital axis of the
binary as the vertical axis. Then, we define the disk surface based on
the heights at which the density decreases by a factor of 0.01 from the
equatorial plane.  Finally, we obtain disk quantities through vertical
integration or averaging between the disk surfaces, depending on the
variable.

The column density shown in the top panel agrees with the 2D density
distribution in Fig.~\ref{fig:evolution_faceon_2d} (center). It exhibits
a peak around $700-1,000\,\mr{au}$, attributed to the circum-stellar
disks, a valley between $1,000-2,000\,\mr{au}$ resulting from the
binary-induced gaps, and a gradually decreasing slope outside,
corresponding to the circum-binary disk.

The second panel displays the temperature, indicating that the gas
within the circum-binary disk is roughly isothermal. Enhancement near
the protostars is likely a result of the radiative feedback from them.

In the third panel, we plot the rotational velocity, together with the
two Keplerian velocities, considering only the mass of the protostars
and considering both the mass of the protostars and the enclosed gas
mass.  Except for $r\sim1,500\,\mr{au}$, where the binary's perturbation
is significant, the rotational velocity lies between the two Keplerian
velocities. This suggests that the gas self-gravity and pressure
gradient play substantial roles in the radial force balance within the
circum-binary disk.

The fourth panel illustrates the disk height, $H_\mr{disk}$, and the
scale height, $H_\mr{scale}=c_\mr{s}/\Omega$.  The high pressure
contributing to the radial force balance also influences the vertical
force balance, causing the disk to inflate to an aspect ratio greater
than unity.  Note that the two heights satisfy the relation
$H_\mr{disk}\approx 3\,H_\mr{scale}$ for the hydrostatic equilibrium:
the heights at which the density decreases by a factor of $0.01$ is
about $3\,H_\mr{scale}$ in the hydrostatic profile $\propto
e^{-z^2/2\,H_\mr{scale}^2}$.

In the last panel, Toomre's Q parameter has a marginally stable value of
$Q\approx 1$ across a wide range of the circum-binary disk. This
quantitatively confirms our view in Section~\ref{sec:int} that the
accretion within the circum-binary disk is driven by the self-regulated
gravitational instability.

 \section{Structure on the scale of a circum-stellar disk}
 \label{sec:csd}

 \begin{figure}
  \centering \includegraphics[width=6.3cm]{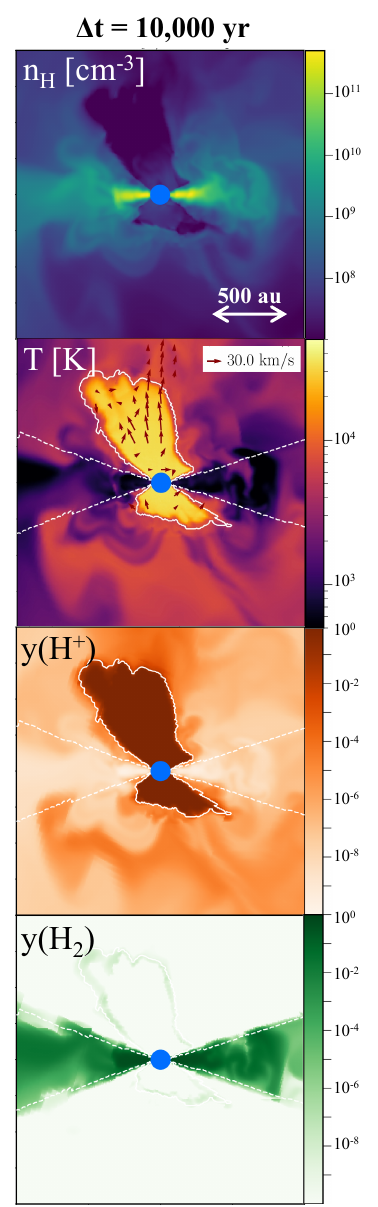}
  \caption{The edge-on slice snapshots for one of the circum-stellar
  disks at $\Delta t = 10,000\,\mr{yr}$ in the Intermediate-$\dot{M}$
  case. From the top to bottom, we plot the density, temperature,
  $\mr{H_2}$ fraction, and $\mr{H^+}$ fraction. In the lower three
  panels, the bipolar photoionization and photodissociation fronts are
  demarcated with solid and dashed lines. }
  \label{fig:circum_stellar_2d}
 \end{figure}

 \begin{figure}
  \centering \includegraphics[width=7cm]{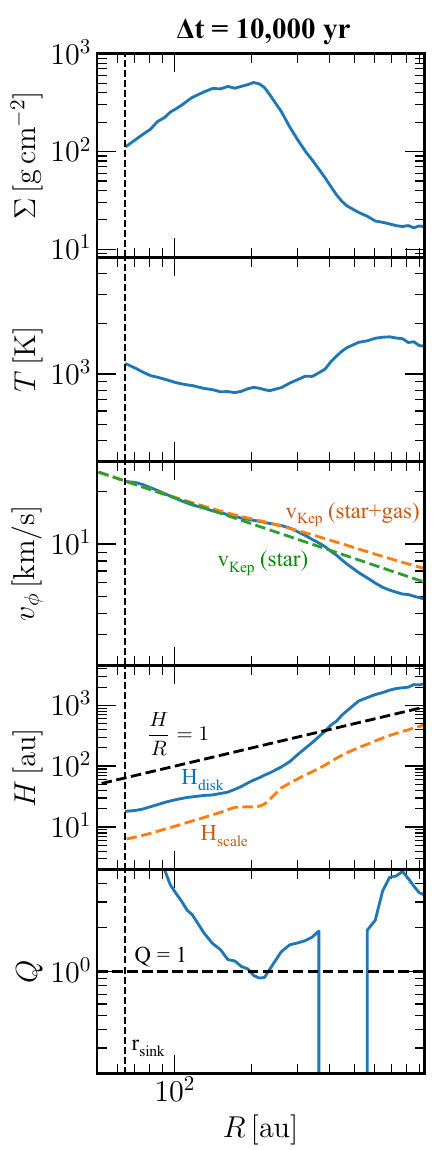}
  \caption{Same as Fig.~\ref{fig:disk1d_cbd} but for the circum-stellar
  disk presented in Fig.~\ref{fig:circum_stellar_2d}. The sink radius is
  indicated with vertical dashed line.}  \label{fig:disk1d_csd1}
 \end{figure}

Here, we examine the binary accretion flow on the scale of
circum-stellar disks. We present the 2D gas distributions in
Fig.~\ref{fig:circum_stellar_2d} and the corresponding 1D profiles in
Fig.~\ref{fig:disk1d_csd1}.

In Fig.~\ref{fig:circum_stellar_2d}, we present the 2D snapshots around
one of the circum-stellar disks inside the circum-binary disk shown in
Figs.~\ref{fig:evolution_faceon_2d} (center),
\ref{fig:evolution_edgeon_2d} (center), \ref{fig:circumbin_2d}, and
\ref{fig:disk1d_cbd}.  The density (top), temperature (second),
$\mr{H_2}$ fraction (third), and $\mr{H^+}$ fraction (last) all indicate
that the gas is accreted onto the central protostar through the
circum-stellar disk while radiative feedback producing bipolar
photoionized and photodissociated regions above and below the disk,
which agrees with the (quasi-)axisymmetric case studied in the
literature (e.g.,
\citealt{Hosokawa:2011aa};\citetalias{Hosokawa:2016aa}).  The gas supply
mechanism to the disk differs from the axisymmetric case, as the gas
enters from one side through a bridge structure, as observed in
Fig.~\ref{fig:evolution_faceon_2d} (center) and explained in
Section~\ref{sec:int}.  However, this difference does not significantly
affect the flow structure on the scale of the circum-stellar disk.
Therefore, the knowledge gained from single Pop III star formation
remains applicable in the context of binary accretion from the
circum-binary disk.

Fig.~\ref{fig:disk1d_csd1} presents the 1D profiles of the
circum-stellar disk illustrated in Fig.~\ref{fig:circum_stellar_2d} in
the same way as in Fig.~\ref{fig:disk1d_cbd}.  The surface density
(top), temperature (second), rotational velocity (third), disk height
(fourth), Toomre's Q parameter (last) all agree with our understanding
of disk accretion under radiative feedback, as described above.
However, at this particular time ($\Delta t = 10,000\,\mr{yr}$), the
disk is relatively small, restricting the radial range unaffected by the
sink sphere ($r_\mr{sink} = 64\,\mr{au}$) and the outer edge of the
circum-stellar disk ($r_\mr{disk} \sim 300\,\mr{au}$) to a region near
$r \sim 200\,\mr{au}$, where Toomre's Q parameter is approximately
unity.  In the proximity of the sink radius, the surface density
decreases due to the influence of the sink particle.  Near the outer
edge of the disk, the pressure becomes significant, and thus the
rotational velocity decreases below the Keplerian value, and the disk
aspect ratio exceeds unity.  We confirm that, at a later time ($\Delta t
= 18,000\,\mr{yr}$), the disk size increases to $r_\mr{disk} \sim
800\,\mr{au}$ (see Fig.~\ref{fig:mini3}, below). Consequently, a finite
range of the disk is marginally stable with $Q \sim 1$, unaffected by
the aforementioned effects.


\begin{thebibliography}{}
\expandafter\ifx\csname natexlab\endcsname\relax\def\natexlab#1{#1}\fi

\bibitem[{{Abbott} {et~al.}(2016){Abbott}, {Abbott}, {Abbott}, {Abernathy},
  {Acernese}, {Ackley}, {Adams}, {Adams}, {Addesso}, {Adhikari}, \&
  et~al.}]{Abbott:2016aa}
{Abbott}, B.~P., {Abbott}, R., {Abbott}, T.~D., {et~al.} 2016, Physical Review
  Letters, 116, 061102

\bibitem[{{Abe} {et~al.}(2021){Abe}, {Yajima}, {Khochfar}, {Dalla Vecchia}, \&
  {Omukai}}]{Abe:2021vp}
{Abe}, M., {Yajima}, H., {Khochfar}, S., {Dalla Vecchia}, C., \& {Omukai}, K.
  2021, \mnras, 508, 3226

\bibitem[{{Abel} {et~al.}(1997){Abel}, {Anninos}, {Zhang}, \&
  {Norman}}]{Abel:1997aa}
{Abel}, T., {Anninos}, P., {Zhang}, Y., \& {Norman}, M.~L. 1997, New Astron.,
  2, 181

\bibitem[{{Abel} {et~al.}(2002){Abel}, {Bryan}, \& {Norman}}]{Abel:2002aa}
{Abel}, T., {Bryan}, G.~L., \& {Norman}, M.~L. 2002, Science, 295, 93

\bibitem[{{Abel} \& {Wandelt}(2002)}]{Abel:2002ab}
{Abel}, T., \& {Wandelt}, B.~D. 2002, \mnras, 330, L53

\bibitem[{{Alvarez} {et~al.}(2009){Alvarez}, {Wise}, \&
  {Abel}}]{Alvarez:2009aa}
{Alvarez}, M.~A., {Wise}, J.~H., \& {Abel}, T. 2009, \apjl, 701, L133

\bibitem[{{Anninos} {et~al.}(1997){Anninos}, {Zhang}, {Abel}, \&
  {Norman}}]{Anninos:1997aa}
{Anninos}, P., {Zhang}, Y., {Abel}, T., \& {Norman}, M.~L. 1997, New Astron.,
  2, 209

\bibitem[{{Bate} \& {Bonnell}(1997)}]{Bate:1997ab}
{Bate}, M.~R., \& {Bonnell}, I.~A. 1997, \mnras, 285, 33

\bibitem[{{Belczynski} {et~al.}(2017){Belczynski}, {Ryu}, {Perna}, {Berti},
  {Tanaka}, \& {Bulik}}]{Belczynski:2017aa}
{Belczynski}, K., {Ryu}, T., {Perna}, R., {et~al.} 2017, \mnras, 471, 4702

\bibitem[{{Berger} \& {Colella}(1989)}]{Berger:1989vo}
{Berger}, M.~J., \& {Colella}, P. 1989, Journal of Computational Physics, 82,
  64

\bibitem[{{Berger} \& {Oliger}(1984)}]{Berger:1984ui}
{Berger}, M.~J., \& {Oliger}, J. 1984, Journal of Computational Physics, 53,
  484

\bibitem[{{Bromm} {et~al.}(2002){Bromm}, {Coppi}, \& {Larson}}]{Bromm:2002aa}
{Bromm}, V., {Coppi}, P.~S., \& {Larson}, R.~B. 2002, \apj, 564, 23

\bibitem[{{Chiaki} {et~al.}(2018){Chiaki}, {Susa}, \& {Hirano}}]{Chiaki:2018ab}
{Chiaki}, G., {Susa}, H., \& {Hirano}, S. 2018, \mnras, 475, 4378

\bibitem[{{Chon} \& {Hosokawa}(2019)}]{Chon:2019aa}
{Chon}, S., \& {Hosokawa}, T. 2019, \mnras, 488, 2658

\bibitem[{{Clark} {et~al.}(2011){Clark}, {Glover}, {Smith}, {Greif}, {Klessen},
  \& {Bromm}}]{Clark:2011ab}
{Clark}, P.~C., {Glover}, S.~C.~O., {Smith}, R.~J., {et~al.} 2011, Science,
  331, 1040

\bibitem[{{Couchman} \& {Rees}(1986)}]{Couchman:1986aa}
{Couchman}, H.~M.~P., \& {Rees}, M.~J. 1986, \mnras, 221, 53

\bibitem[{{Dewdney} {et~al.}(2009){Dewdney}, {Hall}, {Schilizzi}, \&
  {Lazio}}]{Dewdney:2009aa}
{Dewdney}, P.~E., {Hall}, P.~J., {Schilizzi}, R.~T., \& {Lazio}, T.~J.~L.~W.
  2009, IEEE Proceedings, 97, 1482

\bibitem[{{Draine} \& {Bertoldi}(1996)}]{Draine:1996aa}
{Draine}, B.~T., \& {Bertoldi}, F. 1996, \apj, 468, 269

\bibitem[{{Duch{\^e}ne} \& {Kraus}(2013)}]{Duchene:2013aa}
{Duch{\^e}ne}, G., \& {Kraus}, A. 2013, \araa, 51, 269

\bibitem[{{Federrath} {et~al.}(2010){Federrath}, {Banerjee}, {Clark}, \&
  {Klessen}}]{Federrath:2010ab}
{Federrath}, C., {Banerjee}, R., {Clark}, P.~C., \& {Klessen}, R.~S. 2010,
  \apj, 713, 269

\bibitem[{{Federrath} {et~al.}(2011){Federrath}, {Sur}, {Schleicher},
  {Banerjee}, \& {Klessen}}]{Federrath:2011aa}
{Federrath}, C., {Sur}, S., {Schleicher}, D.~R.~G., {Banerjee}, R., \&
  {Klessen}, R.~S. 2011, \apj, 731, 62

\bibitem[{{Forrey}(2013)}]{Forrey:2013aa}
{Forrey}, R.~C. 2013, \apjl, 773, L25

\bibitem[{{Fukushima} {et~al.}(2018){Fukushima}, {Omukai}, \&
  {Hosokawa}}]{Fukushima:2018aa}
{Fukushima}, H., {Omukai}, K., \& {Hosokawa}, T. 2018, \mnras, 473, 4754

\bibitem[{{Galli} \& {Palla}(1998)}]{Galli:1998aa}
{Galli}, D., \& {Palla}, F. 1998, \aap, 335, 403

\bibitem[{{Garcia} {et~al.}(2023){Garcia}, {Ricotti}, {Sugimura}, \&
  {Park}}]{Garcia:2023tg}
{Garcia}, F. A.~B., {Ricotti}, M., {Sugimura}, K., \& {Park}, J. 2023, \mnras,
  522, 2495

\bibitem[{{Glover}(2013)}]{Glover:2013aa}
{Glover}, S. 2013, Astrophysics and Space Science Library, Vol. 396, {The First
  Stars}, ed. T.~{Wiklind}, B.~{Mobasher}, \& V.~{Bromm}, 103

\bibitem[{{Glover}(2015)}]{Glover:2015ab}
{Glover}, S.~C.~O. 2015, \mnras, 453, 2901

\bibitem[{{Glover} \& {Jappsen}(2007)}]{Glover:2007aa}
{Glover}, S.~C.~O., \& {Jappsen}, A.-K. 2007, \apj, 666, 1

\bibitem[{{G{\'o}rski} {et~al.}(2005){G{\'o}rski}, {Hivon}, {Banday},
  {Wandelt}, {Hansen}, {Reinecke}, \& {Bartelmann}}]{Gorski:2005aa}
{G{\'o}rski}, K.~M., {Hivon}, E., {Banday}, A.~J., {et~al.} 2005, \apj, 622,
  759

\bibitem[{{Greif}(2015)}]{Greif:2015aa}
{Greif}, T.~H. 2015, Computational Astrophysics and Cosmology, 2, 3

\bibitem[{{Greif} {et~al.}(2012){Greif}, {Bromm}, {Clark}, {Glover}, {Smith},
  {Klessen}, {Yoshida}, \& {Springel}}]{Greif:2012aa}
{Greif}, T.~H., {Bromm}, V., {Clark}, P.~C., {et~al.} 2012, \mnras, 424, 399

\bibitem[{{Greif} {et~al.}(2011){Greif}, {Springel}, {White}, {Glover},
  {Clark}, {Smith}, {Klessen}, \& {Bromm}}]{Greif:2011aa}
{Greif}, T.~H., {Springel}, V., {White}, S.~D.~M., {et~al.} 2011, \apj, 737, 75

\bibitem[{{Hartwig} {et~al.}(2015){Hartwig}, {Bromm}, {Klessen}, \&
  {Glover}}]{Hartwig:2015ac}
{Hartwig}, T., {Bromm}, V., {Klessen}, R.~S., \& {Glover}, S.~C.~O. 2015,
  \mnras, 447, 3892

\bibitem[{{Hartwig} {et~al.}(2016){Hartwig}, {Volonteri}, {Bromm}, {Klessen},
  {Barausse}, {Magg}, \& {Stacy}}]{Hartwig:2016aa}
{Hartwig}, T., {Volonteri}, M., {Bromm}, V., {et~al.} 2016, \mnras, 460, L74

\bibitem[{{Haworth} {et~al.}(2017){Haworth}, {Facchini}, {Clarke}, \&
  {Cleeves}}]{Haworth:2017tj}
{Haworth}, T.~J., {Facchini}, S., {Clarke}, C.~J., \& {Cleeves}, L.~I. 2017,
  \mnras, 468, L108

\bibitem[{{Heath} \& {Nixon}(2020)}]{Heath:2020aa}
{Heath}, R.~M., \& {Nixon}, C.~J. 2020, \aap, 641, A64

\bibitem[{{Higashi} {et~al.}(2021){Higashi}, {Susa}, \&
  {Chiaki}}]{Higashi:2021vt}
{Higashi}, S., {Susa}, H., \& {Chiaki}, G. 2021, \apj, 915, 107

\bibitem[{{Higashi} {et~al.}(2022){Higashi}, {Susa}, \&
  {Chiaki}}]{Higashi:2022up}
---. 2022, \apj, 940, 38

\bibitem[{{Hirano} \& {Bromm}(2017)}]{Hirano:2017aa}
{Hirano}, S., \& {Bromm}, V. 2017, \mnras, 470, 898

\bibitem[{{Hirano} {et~al.}(2015){Hirano}, {Hosokawa}, {Yoshida}, {Omukai}, \&
  {Yorke}}]{Hirano:2015aa}
{Hirano}, S., {Hosokawa}, T., {Yoshida}, N., {Omukai}, K., \& {Yorke}, H.~W.
  2015, \mnras, 448, 568

\bibitem[{{Hirano} {et~al.}(2014){Hirano}, {Hosokawa}, {Yoshida}, {Umeda},
  {Omukai}, {Chiaki}, \& {Yorke}}]{Hirano:2014aa}
{Hirano}, S., {Hosokawa}, T., {Yoshida}, N., {et~al.} 2014, \apj, 781, 60

\bibitem[{{Hirano} \& {Machida}(2022)}]{Hirano:2022vs}
{Hirano}, S., \& {Machida}, M.~N. 2022, \apjl, 935, L16

\bibitem[{{Hollenbach} \& {McKee}(1979)}]{Hollenbach:1979aa}
{Hollenbach}, D., \& {McKee}, C.~F. 1979, \apjs, 41, 555

\bibitem[{{Hosokawa} {et~al.}(2016){Hosokawa}, {Hirano}, {Kuiper}, {Yorke},
  {Omukai}, \& {Yoshida}}]{Hosokawa:2016aa}
{Hosokawa}, T., {Hirano}, S., {Kuiper}, R., {et~al.} 2016, \apj, 824, 119

\bibitem[{{Hosokawa} \& {Omukai}(2009)}]{Hosokawa:2009aa}
{Hosokawa}, T., \& {Omukai}, K. 2009, \apj, 691, 823

\bibitem[{{Hosokawa} {et~al.}(2011){Hosokawa}, {Omukai}, {Yoshida}, \&
  {Yorke}}]{Hosokawa:2011aa}
{Hosokawa}, T., {Omukai}, K., {Yoshida}, N., \& {Yorke}, H.~W. 2011, Science,
  334, 1250

\bibitem[{{Hosokawa} {et~al.}(2010){Hosokawa}, {Yorke}, \&
  {Omukai}}]{Hosokawa:2010aa}
{Hosokawa}, T., {Yorke}, H.~W., \& {Omukai}, K. 2010, \apj, 721, 478

\bibitem[{{Hummel} {et~al.}(2015){Hummel}, {Stacy}, {Jeon}, {Oliveri}, \&
  {Bromm}}]{Hummel:2015aa}
{Hummel}, J.~A., {Stacy}, A., {Jeon}, M., {Oliveri}, A., \& {Bromm}, V. 2015,
  \mnras, 453, 4136

\bibitem[{{Ishiyama} {et~al.}(2016){Ishiyama}, {Sudo}, {Yokoi}, {Hasegawa},
  {Tominaga}, \& {Susa}}]{Ishiyama:2016aa}
{Ishiyama}, T., {Sudo}, K., {Yokoi}, S., {et~al.} 2016, \apj, 826, 9

\bibitem[{{Jaura} {et~al.}(2022){Jaura}, {Glover}, {Wollenberg}, {Klessen},
  {Geen}, \& {Haemmerl{\'e}}}]{Jaura:2022vn}
{Jaura}, O., {Glover}, S. C.~O., {Wollenberg}, K. M.~J., {et~al.} 2022, \mnras,
  512, 116

\bibitem[{{Jeon} {et~al.}(2012){Jeon}, {Pawlik}, {Greif}, {Glover}, {Bromm},
  {Milosavljevi{\'c}}, \& {Klessen}}]{Jeon:2012aa}
{Jeon}, M., {Pawlik}, A.~H., {Greif}, T.~H., {et~al.} 2012, \apj, 754, 34

\bibitem[{{John}(1988)}]{John:1988aa}
{John}, T.~L. 1988, \aap, 193, 189

\bibitem[{{Kim} {et~al.}(2017){Kim}, {Kim}, {Ostriker}, \&
  {Skinner}}]{Kim:2017aa}
{Kim}, J.-G., {Kim}, W.-T., {Ostriker}, E.~C., \& {Skinner}, M.~A. 2017, \apj,
  851, 93

\bibitem[{{Kimura} {et~al.}(2021){Kimura}, {Hosokawa}, \&
  {Sugimura}}]{Kimura:2021aa}
{Kimura}, K., {Hosokawa}, T., \& {Sugimura}, K. 2021, \apj, 911, 52

\bibitem[{{Kimura} {et~al.}(2023){Kimura}, {Hosokawa}, {Sugimura}, \&
  {Fukushima}}]{Kimura:2023}
{Kimura}, K., {Hosokawa}, T., {Sugimura}, K., \& {Fukushima}, H. 2023, \apj,
  950, 184

\bibitem[{{Kinugawa} {et~al.}(2014){Kinugawa}, {Inayoshi}, {Hotokezaka},
  {Nakauchi}, \& {Nakamura}}]{Kinugawa:2014aa}
{Kinugawa}, T., {Inayoshi}, K., {Hotokezaka}, K., {Nakauchi}, D., \&
  {Nakamura}, T. 2014, \mnras, 442, 2963

\bibitem[{{Kirihara} {et~al.}(2023){Kirihara}, {Susa}, {Hosokawa}, \&
  {Kinugawa}}]{Kirihara:2023}
{Kirihara}, T., {Susa}, H., {Hosokawa}, T., \& {Kinugawa}, T. 2023, \apj, 950,
  188

\bibitem[{{Klessen} \& {Glover}(2023)}]{Klessen:2023ui}
{Klessen}, R.~S., \& {Glover}, S. C.~O. 2023, arXiv e-prints, arXiv:2303.12500

\bibitem[{{Kreckel} {et~al.}(2010){Kreckel}, {Bruhns}, {{\v C}{\'{\i}}{\v
  z}ek}, {Glover}, {Miller}, {Urbain}, \& {Savin}}]{Kreckel:2010aa}
{Kreckel}, H., {Bruhns}, H., {{\v C}{\'{\i}}{\v z}ek}, M., {et~al.} 2010,
  Science, 329, 69

\bibitem[{{Krumholz} {et~al.}(2007){Krumholz}, {Stone}, \&
  {Gardiner}}]{Krumholz:2007ab}
{Krumholz}, M.~R., {Stone}, J.~M., \& {Gardiner}, T.~A. 2007, \apj, 671, 518

\bibitem[{{Latif} {et~al.}(2022){Latif}, {Whalen}, \&
  {Khochfar}}]{Latif:2022wf}
{Latif}, M.~A., {Whalen}, D., \& {Khochfar}, S. 2022, \apj, 925, 28

\bibitem[{{Liu} \& {Bromm}(2020)}]{Liu:2020a}
{Liu}, B., \& {Bromm}, V. 2020, \apjl, 903, L40

\bibitem[{{Liu} \& {Bromm}(2021)}]{Liu:2021a}
---. 2021, \mnras, 506, 5451

\bibitem[{{Liu} {et~al.}(2021){Liu}, {Meynet}, \& {Bromm}}]{Liu:2021}
{Liu}, B., {Meynet}, G., \& {Bromm}, V. 2021, \mnras, 501, 643

\bibitem[{{Machida} {et~al.}(2008{\natexlab{a}}){Machida}, {Matsumoto}, \&
  {Inutsuka}}]{Machida:2008ab}
{Machida}, M.~N., {Matsumoto}, T., \& {Inutsuka}, S.-i. 2008{\natexlab{a}},
  \apj, 685, 690

\bibitem[{{Machida} {et~al.}(2006){Machida}, {Omukai}, {Matsumoto}, \&
  {Inutsuka}}]{Machida:2006aa}
{Machida}, M.~N., {Omukai}, K., {Matsumoto}, T., \& {Inutsuka}, S.-i. 2006,
  \apjl, 647, L1

\bibitem[{{Machida} {et~al.}(2008{\natexlab{b}}){Machida}, {Omukai},
  {Matsumoto}, \& {Inutsuka}}]{Machida:2008aa}
---. 2008{\natexlab{b}}, \apj, 677, 813

\bibitem[{{Martin} {et~al.}(1996){Martin}, {Schwarz}, \&
  {Mandy}}]{Martin:1996aa}
{Martin}, P.~G., {Schwarz}, D.~H., \& {Mandy}, M.~E. 1996, \apj, 461, 265

\bibitem[{{Matsukoba} {et~al.}(2021){Matsukoba}, {Vorobyov}, {Sugimura},
  {Chon}, {Hosokawa}, \& {Omukai}}]{Matsukoba:2021tb}
{Matsukoba}, R., {Vorobyov}, E.~I., {Sugimura}, K., {et~al.} 2021, \mnras, 500,
  4126

\bibitem[{{Matsumoto}(2007)}]{Matsumoto:2007aa}
{Matsumoto}, T. 2007, \pasj, 59, 905

\bibitem[{{Matsumoto} {et~al.}(2015){Matsumoto}, {Dobashi}, \&
  {Shimoikura}}]{Matsumoto:2015ab}
{Matsumoto}, T., {Dobashi}, K., \& {Shimoikura}, T. 2015, \apj, 801, 77

\bibitem[{{Matsumoto} {et~al.}(2019){Matsumoto}, {Saigo}, \&
  {Takakuwa}}]{Matsumoto:2019wj}
{Matsumoto}, T., {Saigo}, K., \& {Takakuwa}, S. 2019, \apj, 871, 36

\bibitem[{{McKee} {et~al.}(2020){McKee}, {Stacy}, \& {Li}}]{McKee:2020ve}
{McKee}, C.~F., {Stacy}, A., \& {Li}, P.~S. 2020, \mnras, 496, 5528

\bibitem[{{McKee} \& {Tan}(2008)}]{McKee:2008aa}
{McKee}, C.~F., \& {Tan}, J.~C. 2008, \apj, 681, 771

\bibitem[{{Millar}(1991)}]{Millar:1991wg}
{Millar}, T.~J. 1991, \aap, 242, 241

\bibitem[{{Mirabel} {et~al.}(2011){Mirabel}, {Dijkstra}, {Laurent}, {Loeb}, \&
  {Pritchard}}]{Mirabel:2011aa}
{Mirabel}, I.~F., {Dijkstra}, M., {Laurent}, P., {Loeb}, A., \& {Pritchard},
  J.~R. 2011, \aap, 528, A149

\bibitem[{{Moody} {et~al.}(2019){Moody}, {Shi}, \& {Stone}}]{Moody:2019aa}
{Moody}, M. S.~L., {Shi}, J.-M., \& {Stone}, J.~M. 2019, \apj, 875, 66

\bibitem[{{Mu{\~n}oz} {et~al.}(2019){Mu{\~n}oz}, {Miranda}, \&
  {Lai}}]{Munoz:2019aa}
{Mu{\~n}oz}, D.~J., {Miranda}, R., \& {Lai}, D. 2019, \apj, 871, 84

\bibitem[{{Omukai}(2001)}]{Omukai:2001aa}
{Omukai}, K. 2001, \apj, 546, 635

\bibitem[{{Omukai} \& {Inutsuka}(2002)}]{Omukai:2002aa}
{Omukai}, K., \& {Inutsuka}, S.-i. 2002, \mnras, 332, 59

\bibitem[{{Omukai} \& {Nishi}(1998)}]{Omukai:1998aa}
{Omukai}, K., \& {Nishi}, R. 1998, \apj, 508, 141

\bibitem[{{Omukai} \& {Palla}(2003)}]{Omukai:2003ab}
{Omukai}, K., \& {Palla}, F. 2003, \apj, 589, 677

\bibitem[{{Osterbrock} \& {Ferland}(2006)}]{Osterbrock:2006aa}
{Osterbrock}, D.~E., \& {Ferland}, G.~J. 2006, {Astrophysics of gaseous nebulae
  and active galactic nuclei}

\bibitem[{{Palla} {et~al.}(1983){Palla}, {Salpeter}, \&
  {Stahler}}]{Palla:1983vg}
{Palla}, F., {Salpeter}, E.~E., \& {Stahler}, S.~W. 1983, \apj, 271, 632

\bibitem[{{Park} {et~al.}(2021{\natexlab{a}}){Park}, {Ricotti}, \&
  {Sugimura}}]{Park:2021aa}
{Park}, J., {Ricotti}, M., \& {Sugimura}, K. 2021{\natexlab{a}}, \mnras, 508,
  6176

\bibitem[{{Park} {et~al.}(2021{\natexlab{b}}){Park}, {Ricotti}, \&
  {Sugimura}}]{Park:2021ab}
---. 2021{\natexlab{b}}, \mnras, 508, 6193

\bibitem[{{Park} {et~al.}(2023){Park}, {Ricotti}, \& {Sugimura}}]{Park:2023ta}
---. 2023, \mnras, 521, 5334

\bibitem[{{Park} \& {Bogdanovi{\'c}}(2017)}]{Park:2017aa}
{Park}, K., \& {Bogdanovi{\'c}}, T. 2017, \apj, 838, 103

\bibitem[{{Prole} {et~al.}(2022{\natexlab{a}}){Prole}, {Clark}, {Klessen}, \&
  {Glover}}]{Prole:2022uh}
{Prole}, L.~R., {Clark}, P.~C., {Klessen}, R.~S., \& {Glover}, S. C.~O.
  2022{\natexlab{a}}, \mnras, 510, 4019

\bibitem[{{Prole} {et~al.}(2022{\natexlab{b}}){Prole}, {Clark}, {Klessen},
  {Glover}, \& {Pakmor}}]{Prole:2022tq}
{Prole}, L.~R., {Clark}, P.~C., {Klessen}, R.~S., {Glover}, S. C.~O., \&
  {Pakmor}, R. 2022{\natexlab{b}}, \mnras, 516, 2223

\bibitem[{{Ricotti}(2016)}]{Ricotti:2016ab}
{Ricotti}, M. 2016, \mnras, 462, 601

\bibitem[{{Ricotti} {et~al.}(2016){Ricotti}, {Parry}, \&
  {Gnedin}}]{Ricotti:2016aa}
{Ricotti}, M., {Parry}, O.~H., \& {Gnedin}, N.~Y. 2016, \apj, 831, 204

\bibitem[{{Rosen} {et~al.}(2017){Rosen}, {Krumholz}, {Oishi}, {Lee}, \&
  {Klein}}]{Rosen:2017aa}
{Rosen}, A.~L., {Krumholz}, M.~R., {Oishi}, J.~S., {Lee}, A.~T., \& {Klein},
  R.~I. 2017, Journal of Computational Physics, 330, 924

\bibitem[{{Saad} {et~al.}(2022){Saad}, {Bromm}, \& {El Eid}}]{Saad:2022vg}
{Saad}, C.~R., {Bromm}, V., \& {El Eid}, M. 2022, \mnras, 516, 3130

\bibitem[{{Sadanari} {et~al.}(2021){Sadanari}, {Omukai}, {Sugimura},
  {Matsumoto}, \& {Tomida}}]{Sadanari:2021aa}
{Sadanari}, K.~E., {Omukai}, K., {Sugimura}, K., {Matsumoto}, T., \& {Tomida},
  K. 2021, \mnras, 505, 4197

\bibitem[{{Sadanari} {et~al.}(2023){Sadanari}, {Omukai}, {Sugimura},
  {Matsumoto}, \& {Tomida}}]{Sadanari:2023vd}
---. 2023, \mnras, 519, 3076

\bibitem[{{Satsuka} {et~al.}(2017){Satsuka}, {Tsuribe}, {Tanaka}, \&
  {Nagamine}}]{Satsuka:2017aa}
{Satsuka}, T., {Tsuribe}, T., {Tanaka}, S., \& {Nagamine}, K. 2017, \mnras,
  465, 986

\bibitem[{{Schaerer}(2002)}]{Schaerer:2002aa}
{Schaerer}, D. 2002, \aap, 382, 28

\bibitem[{{Schleicher} {et~al.}(2010){Schleicher}, {Banerjee}, {Sur},
  {Arshakian}, {Klessen}, {Beck}, \& {Spaans}}]{Schleicher:2010}
{Schleicher}, D.~R.~G., {Banerjee}, R., {Sur}, S., {et~al.} 2010, \aap, 522,
  A115

\bibitem[{{Schober} {et~al.}(2012){Schober}, {Schleicher}, {Federrath},
  {Glover}, {Klessen}, \& {Banerjee}}]{Schober:2012tb}
{Schober}, J., {Schleicher}, D., {Federrath}, C., {et~al.} 2012, \apj, 754, 99

\bibitem[{{Shapiro} \& {Kang}(1987)}]{Shapiro:1987aa}
{Shapiro}, P.~R., \& {Kang}, H. 1987, \apj, 318, 32

\bibitem[{{Sharda} {et~al.}(2020){Sharda}, {Federrath}, \&
  {Krumholz}}]{Sharda:2020vd}
{Sharda}, P., {Federrath}, C., \& {Krumholz}, M.~R. 2020, \mnras, 497, 336

\bibitem[{{Sharda} {et~al.}(2021){Sharda}, {Federrath}, {Krumholz}, \&
  {Schleicher}}]{Sharda:2021tc}
{Sharda}, P., {Federrath}, C., {Krumholz}, M.~R., \& {Schleicher}, D. R.~G.
  2021, \mnras, 503, 2014

\bibitem[{{Sharda} {et~al.}(2019){Sharda}, {Krumholz}, \&
  {Federrath}}]{Sharda:2019aa}
{Sharda}, P., {Krumholz}, M.~R., \& {Federrath}, C. 2019, \mnras, 490, 513

\bibitem[{{Smith} {et~al.}(2011){Smith}, {Glover}, {Clark}, {Greif}, \&
  {Klessen}}]{Smith:2011aa}
{Smith}, R.~J., {Glover}, S.~C.~O., {Clark}, P.~C., {Greif}, T., \& {Klessen},
  R.~S. 2011, \mnras, 414, 3633

\bibitem[{{Stacy} \& {Bromm}(2013)}]{Stacy:2013aa}
{Stacy}, A., \& {Bromm}, V. 2013, \mnras, 433, 1094

\bibitem[{{Stacy} {et~al.}(2016){Stacy}, {Bromm}, \& {Lee}}]{Stacy:2016aa}
{Stacy}, A., {Bromm}, V., \& {Lee}, A.~T. 2016, \mnras, 462, 1307

\bibitem[{{Stacy} {et~al.}(2012){Stacy}, {Greif}, \& {Bromm}}]{Stacy:2012aa}
{Stacy}, A., {Greif}, T.~H., \& {Bromm}, V. 2012, \mnras, 422, 290

\bibitem[{{Stacy} {et~al.}(2022){Stacy}, {McKee}, {Lee}, {Klein}, \&
  {Li}}]{Stacy:2022ua}
{Stacy}, A., {McKee}, C.~F., {Lee}, A.~T., {Klein}, R.~I., \& {Li}, P.~S. 2022,
  \mnras, 511, 5042

\bibitem[{{Stahler} {et~al.}(1986){Stahler}, {Palla}, \&
  {Salpeter}}]{Stahler:1986aa}
{Stahler}, S.~W., {Palla}, F., \& {Salpeter}, E.~E. 1986, \apj, 302, 590

\bibitem[{{Sugimura} {et~al.}(2018){Sugimura}, {Hosokawa}, {Yajima},
  {Inayoshi}, \& {Omukai}}]{Sugimura:2018aa}
{Sugimura}, K., {Hosokawa}, T., {Yajima}, H., {Inayoshi}, K., \& {Omukai}, K.
  2018, \mnras, 478, 3961

\bibitem[{{Sugimura} {et~al.}(2017){Sugimura}, {Hosokawa}, {Yajima}, \&
  {Omukai}}]{Sugimura:2017ab}
{Sugimura}, K., {Hosokawa}, T., {Yajima}, H., \& {Omukai}, K. 2017, \mnras,
  469, 62

\bibitem[{{Sugimura} {et~al.}(2020){Sugimura}, {Matsumoto}, {Hosokawa},
  {Hirano}, \& {Omukai}}]{Sugimura:2020aa}
{Sugimura}, K., {Matsumoto}, T., {Hosokawa}, T., {Hirano}, S., \& {Omukai}, K.
  2020, \apjl, 892, L14

\bibitem[{{Sugimura} \& {Ricotti}(2020)}]{Sugimura:2020ab}
{Sugimura}, K., \& {Ricotti}, M. 2020, \mnras, 495, 2966

\bibitem[{{Sur} {et~al.}(2010){Sur}, {Schleicher}, {Banerjee}, {Federrath}, \&
  {Klessen}}]{Sur:2010aa}
{Sur}, S., {Schleicher}, D.~R.~G., {Banerjee}, R., {Federrath}, C., \&
  {Klessen}, R.~S. 2010, \apjl, 721, L134

\bibitem[{{Susa}(2013)}]{Susa:2013aa}
{Susa}, H. 2013, \apj, 773, 185

\bibitem[{{Susa}(2019)}]{Susa:2019wj}
---. 2019, \apj, 877, 99

\bibitem[{{Susa} {et~al.}(2014){Susa}, {Hasegawa}, \& {Tominaga}}]{Susa:2014aa}
{Susa}, H., {Hasegawa}, K., \& {Tominaga}, N. 2014, \apj, 792, 32

\bibitem[{{Tan} \& {McKee}(2004)}]{Tan:2004aa}
{Tan}, J.~C., \& {McKee}, C.~F. 2004, \apj, 603, 383

\bibitem[{{Tanikawa} {et~al.}(2021){Tanikawa}, {Susa}, {Yoshida}, {Trani}, \&
  {Kinugawa}}]{Tanikawa:2021aa}
{Tanikawa}, A., {Susa}, H., {Yoshida}, T., {Trani}, A.~A., \& {Kinugawa}, T.
  2021, \apj, 910, 30

\bibitem[{{Tegmark} {et~al.}(1997){Tegmark}, {Silk}, {Rees}, {Blanchard},
  {Abel}, \& {Palla}}]{Tegmark:1997aa}
{Tegmark}, M., {Silk}, J., {Rees}, M.~J., {et~al.} 1997, \apj, 474, 1

\bibitem[{{Tiede} {et~al.}(2020){Tiede}, {Zrake}, {MacFadyen}, \&
  {Haiman}}]{Tiede:2020aa}
{Tiede}, C., {Zrake}, J., {MacFadyen}, A., \& {Haiman}, Z. 2020, \apj, 900, 43

\bibitem[{{Toyouchi} {et~al.}(2020){Toyouchi}, {Hosokawa}, {Sugimura}, \&
  {Kuiper}}]{Toyouchi:2020uf}
{Toyouchi}, D., {Hosokawa}, T., {Sugimura}, K., \& {Kuiper}, R. 2020, \mnras,
  496, 1909

\bibitem[{{Truelove} {et~al.}(1997){Truelove}, {Klein}, {McKee}, {Holliman},
  {Howell}, \& {Greenough}}]{Truelove:1997aa}
{Truelove}, J.~K., {Klein}, R.~I., {McKee}, C.~F., {et~al.} 1997, \apjl, 489,
  L179

\bibitem[{{Turk} {et~al.}(2012){Turk}, {Oishi}, {Abel}, \&
  {Bryan}}]{Turk:2012aa}
{Turk}, M.~J., {Oishi}, J.~S., {Abel}, T., \& {Bryan}, G.~L. 2012, \apj, 745,
  154

\bibitem[{{Wise} \& {Abel}(2011)}]{Wise:2011aa}
{Wise}, J.~H., \& {Abel}, T. 2011, \mnras, 414, 3458

\bibitem[{{Wise} {et~al.}(2012){Wise}, {Turk}, {Norman}, \&
  {Abel}}]{Wise:2012aa}
{Wise}, J.~H., {Turk}, M.~J., {Norman}, M.~L., \& {Abel}, T. 2012, \apj, 745,
  50

\bibitem[{{Wolcott-Green} \& {Haiman}(2019)}]{Wolcott-Green:2019vl}
{Wolcott-Green}, J., \& {Haiman}, Z. 2019, \mnras, 484, 2467

\bibitem[{{Wollenberg} {et~al.}(2020){Wollenberg}, {Glover}, {Clark}, \&
  {Klessen}}]{Wollenberg:2020wo}
{Wollenberg}, K. M.~J., {Glover}, S. C.~O., {Clark}, P.~C., \& {Klessen}, R.~S.
  2020, \mnras, 494, 1871

\bibitem[{{Woosley} {et~al.}(2002){Woosley}, {Heger}, \&
  {Weaver}}]{Woosley:2002aa}
{Woosley}, S.~E., {Heger}, A., \& {Weaver}, T.~A. 2002, Reviews of Modern
  Physics, 74, 1015

\bibitem[{{Xu} {et~al.}(2008){Xu}, {O'Shea}, {Collins}, {Norman}, {Li}, \&
  {Li}}]{Xu:2008tb}
{Xu}, H., {O'Shea}, B.~W., {Collins}, D.~C., {et~al.} 2008, \apjl, 688, L57

\bibitem[{{Yorke} \& {Welz}(1996)}]{Yorke:1996aa}
{Yorke}, H.~W., \& {Welz}, A. 1996, \aap, 315, 555

\bibitem[{{Yoshida} {et~al.}(2008){Yoshida}, {Omukai}, \&
  {Hernquist}}]{Yoshida:2008aa}
{Yoshida}, N., {Omukai}, K., \& {Hernquist}, L. 2008, Science, 321, 669

\end{thebibliography}
\end{document}